\documentclass[aps,prl,twocolumn,groupedaddress]{revtex4-2}

\usepackage{graphicx}
\usepackage{dcolumn}
\usepackage{bm}
\usepackage{amsmath}
\usepackage{amssymb}
\usepackage{hyperref}
\usepackage{color}
\usepackage{subcaption}
\usepackage[font=small,labelfont=bf]{caption}
\usepackage{ragged2e}
\usepackage{CJKutf8}           

\usepackage[normalem]{ulem}
\usepackage{cancel}

\hypersetup{hypertex=true,
            colorlinks=true,
            linkcolor=blue,
            anchorcolor=blue,
            citecolor=blue}

\newcommand{\isec}[1]{\noindent\hspace{1em}\textit{#1}---}
\newcommand{\newpara}{\\ \indent}

\newcommand{\x}{$x$ }

\newcommand{\z}{$z$ }

\newcommand{\e}{\text{e} }
\newcommand{\bq}{$\textbf{q}$ }

\newcommand{\half}{\frac{1}{2}}
\newcommand{\tb}[1]{\textbf{#1}}
\newcommand{\mP}{\mathcal{P}}
\newcommand{\T}{\mathcal{T}}

\newcommand{\be}{\begin{eqnarray}}
\newcommand{\ee}{\end{eqnarray}}
\newcommand{\figref}[2]{\textup{FIG.~\ref{#1}.#2}}

\newcommand{\Rmnum}[1]{\uppercase\expandafter{\romannumeral #1}}

\begin{document}
\begin{CJK*}{UTF8}{gbsn}

\title{Nonreciprocal Perfect Coulomb Drag in Electron–Hole Bilayers:\\ Coherent Exciton Superflow as a Diode}


\author{Jun-Xiao Hui \CJKfamily{gbsn}{(辉隽骁)}$^1$,  Qing-Dong Jiang \CJKfamily{gbsn}{(蒋庆东)}$^{1,2}$}
\email{qingdong.jiang@sjtu.edu.cn}
\affiliation{$^1$Tsung-Dao Lee Institute \& School of Physics and Astronomy, Shanghai Jiao Tong University, Pudong, Shanghai, 201210, China\\
$^2$Shanghai Branch, Hefei National Laboratory, Shanghai, 201315, China.
}


\begin{abstract} 

Distinguishing an exciton condensate from an excitonic gas or insulator remains a fundamental challenge, as both phases feature bound electron–hole pairs but differ only by the emergence of macroscopic phase coherence. Here, we theoretically propose that a spin–orbit–coupled bilayer system can host a finite-momentum exciton condensate exhibiting a nonreciprocal perfect Coulomb drag—{\it the coherent-exciton diode effect}. This effect arises from the simultaneous breaking of inversion and time-reversal symmetries in the exciton condensate, resulting in direction-dependent critical counterflow currents. The resulting nonreciprocal perfect Coulomb drag provides a clear and unambiguous transport signature of phase-coherent exciton condensation, offering a powerful and experimentally accessible approach to identify, probe, and control exciton superfluidity in solid-state platforms.

\end{abstract}


\maketitle
\end{CJK*}



 \isec{Introduction.}Excitons, first proposed by Frenkel~\cite{Frenkel1931} to explain the light absorption of solids, are bound states of an electron and a hole that appear widely in semiconductors. In the early days, excitons were viewed as hydrogen-like excitations in semiconductors, manifesting themselves as sharp optical absorption peaks superimposed on the broader onset of electron–hole continuum excitations~\cite{Sturge1962optical,Karni2022trARPESexciton}. These optically excited excitons have finite lifetime and eventually decay. When the electromagnetic field is quantized and its full dynamics is included, photons can hybridize with excitons to form mixed light–matter quasiparticles known as exciton-polaritons~\cite{Hopfield1958excitonpolariton}. By continuously pumping photons to maintain a non-equilibrium steady state, one can realize long-lived exciton-polaritons and even make them condensate~\cite{Snoke2002excitonpolariton}.
\newpara Can excitons and their condensates be stabilized in equilibrium? Inspired by the BCS theory of superconductivity 
\cite{BCS1957}, theorists proposed an affirmative answer in the 1960s~\cite{Blatt1962exciton,Keldysh1964exciton,Jerome1967EI}. A BCS-like ground state of excitons is known as the excitonic insulator~\cite{Jerome1967EI}. The essential condition for the spontaneous formation of excitons is that their binding energy exceeds the band gap~\cite{Pereira2022topoexciton,XieMac2018Xresevoir}. However, excitons are charge-neutral and therefore insensitive to conventional charge transport measurements.
With the advent of advanced fabrication techniques for two-dimensional materials at the end of the 20th century, it became possible to engineer spatially separated electron–hole bilayers. This realization opened the door to interlayer excitons, where the electron and hole reside in different layers~\cite{LozovikYudson1976ehSC,Fogler2014BX,FilShevchenko2018ehSCreview,ZhuDasSarma2024bilayer,YangDaiLi2025EIreview}. Numerous efforts have since been made to create exciton condensates in such bilayer systems, with early experiments in the quantum Hall regime providing suggestive evidence for such a phase~\cite{Eisenstein2014excitonQHreview}. More recently, equilibrium exciton fluid has also been realized in layered 2D materials without out-of-plane magnetic fields~\cite{WangMak2019evidenceXC,MaMak2021EIindoublelayer,NguyenShanMak2025XCPCD,NguyenMak2025quantumoscillationEI,Moon2025VdWexcitonreview,QiMacWang20262compXC}, and at low enough temperature proved to be exciton condensate~\cite{QiMacWang20262compXC,WuMacdonald2015theory}.
A major advantage of electron–hole bilayer systems is the ability to probe excitonic states through transport techniques, particularly by Coulomb drag experiments~\cite{Narozhny2016dragreview}. 
In the exciton condensate phase, one expects a dissipationless, perfect Coulomb drag, where a current in one layer induces an equal and opposite current in the other layer—a hallmark of the formation of tightly bound, interlayer excitons~\cite{Nandi2012XCPCD,NguyenShanMak2025XCPCD,QiWang2025PCD}.

However, perfect Coulomb drag alone is insufficient to identify a dissipationless exciton superfluid phase~\cite{QiWang2025PCD}. An important challenge in exciton physics is therefore to unambiguously distinguish between an excitonic insulator, in which electron–hole pairs form without global coherence, and an exciton superfluid, where macroscopic phase coherence enables dissipationless counterflow. This requires transport signatures that directly probe phase stiffness rather than mere pairing. In this Letter, we show that a spin–orbit–coupled bilayer system can host a finite-momentum exciton condensate exhibiting a perfect nonreciprocal Coulomb drag—which we call {\it coherent-exciton diode effect} (CEDE). This nonreciprocal drag requires global phase-coherent superflow, providing a smoking-gun signature of exciton superfluidity.
\begin{figure}[htbp]
\includegraphics[width=0.46\textwidth]{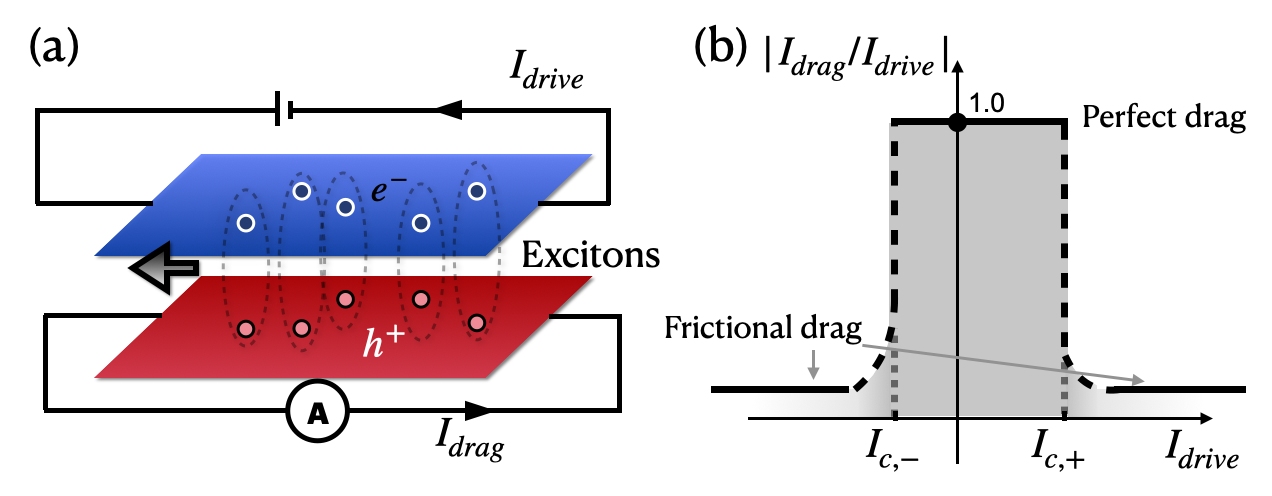}
\caption{\label{fig:dragsketch} \justifying (a) An illustration of Coulomb drag experiment. A current $I_{\rm drive}$ in top layer will spontaneously induce $I_{drag}$ in bottom layer.\, (b) A schematic plot for drag ratio $\zeta=I_{\rm drag}/I_{\rm drive}$: at small $I_{\rm drive}$, $|\zeta|$ is close to 1, corresponding to the perfect Coulomb drag; as $I_{\rm drive}$ exceeds its critical value $I_{c,\pm}$ in either direction, the condensate is broken down, corresponding to the transition to the frictional drag regime between the two Fermi liquids on each layer~\cite{Narozhny2016dragreview}.}
\end{figure}

Our work is inspired by previous seminal studies on supercurrent diode effects in superconductors~\cite{Ando2020SDEobservation,YuanFu2022SDE,HeNagaosa2022SDE,Daido2022SDE,Nadeem2023SDEreview,JiangHu2022SDE}. In recent years, it has been shown that breaking parity ($\mathcal P$) and time-reversal ($\mathcal T$) symmetries in superconductors can not only induce unconventional condensates—such as helical superconductors~\cite{Mineev1994helicalSC}, Fulde–Ferrell–Larkin–Ovchinnikov (FFLO) states~\cite{FF1964,LO1965}, and 
p-wave spin-triplet superconductors~\cite{Volovik2003universeHe}—but also lead to nonreciprocity of the superconducting critical current, known as the supercurrent diode effect. This phenomenon has been observed in both bulk superconductors~\cite{Ando2020SDEobservation,LeLinWu2024KagomeSDE,Lin2022zeroBSDE} and Josephson junctions~\cite{Wu2022zeroBJDE,Trahms2023JDE,ZhangHuJiang2022JDE,HuSunLaw2023josephsondiode}.

We propose that CEDE can arise in asymmetric spin–orbit coupled electron–hole bilayer systems. Experimentally, this may be realized in a sandwich structure, where the top and bottom layers—e.g. transition-metal dichalcogenide materials—have distinct spin–orbit coupling~\cite{MaruyamaSigrist2012sandwich}.
\figref{fig:dragsketch}(a) shows schematics of the proposed setup. An electric current \( I_{\mathrm{drive}} \) which flows through the top layer can induce a spontaneous {drag current} \( I_{\mathrm{drag}} \) in the bottom layer, characterized by the dimensionless drag ratio 
\begin{equation}
    \zeta = \frac{I_{\mathrm{drag}}}{I_{\mathrm{drive}}} .\label{eq:zetaratio}
\end{equation}
In the weakly correlated regime, perturbation theory yields the so-called frictional drag with $\zeta\ll 1$~\cite{Narozhny2016dragreview}. 
In contrast, when interlayer exciton forms, the drag becomes nearly perfect, approaching $\zeta=-1$. 
Previous studies have discussed nonreciprocal frictional drag effects arising from broken spatial inversion $\mP$ and time-reversal $\T$ symmetries, as well as the possibility of nonreciprocal frictional drag~\cite{FuHe2025NRCD,ZverevichLevchenko2025NRCD}. However, the nonreciprocal properties of perfect Coulomb drag remain unexplored. As illustrated in
\figref{fig:dragsketch}(b),
nonreciprocal perfect Coulomb drag manifests itself as a plateau with $|\zeta|=1$, bounded by direction-dependent critical currents $I_{c-}$ and $I_{c+}$. 
When these critical currents differ $I_{c-} \neq  I_{c+}$
the plateau becomes directionally biased, marking the onset of the coherent-exciton diode effect.

\begin{figure}[htbp]
\includegraphics[width=0.46\textwidth]{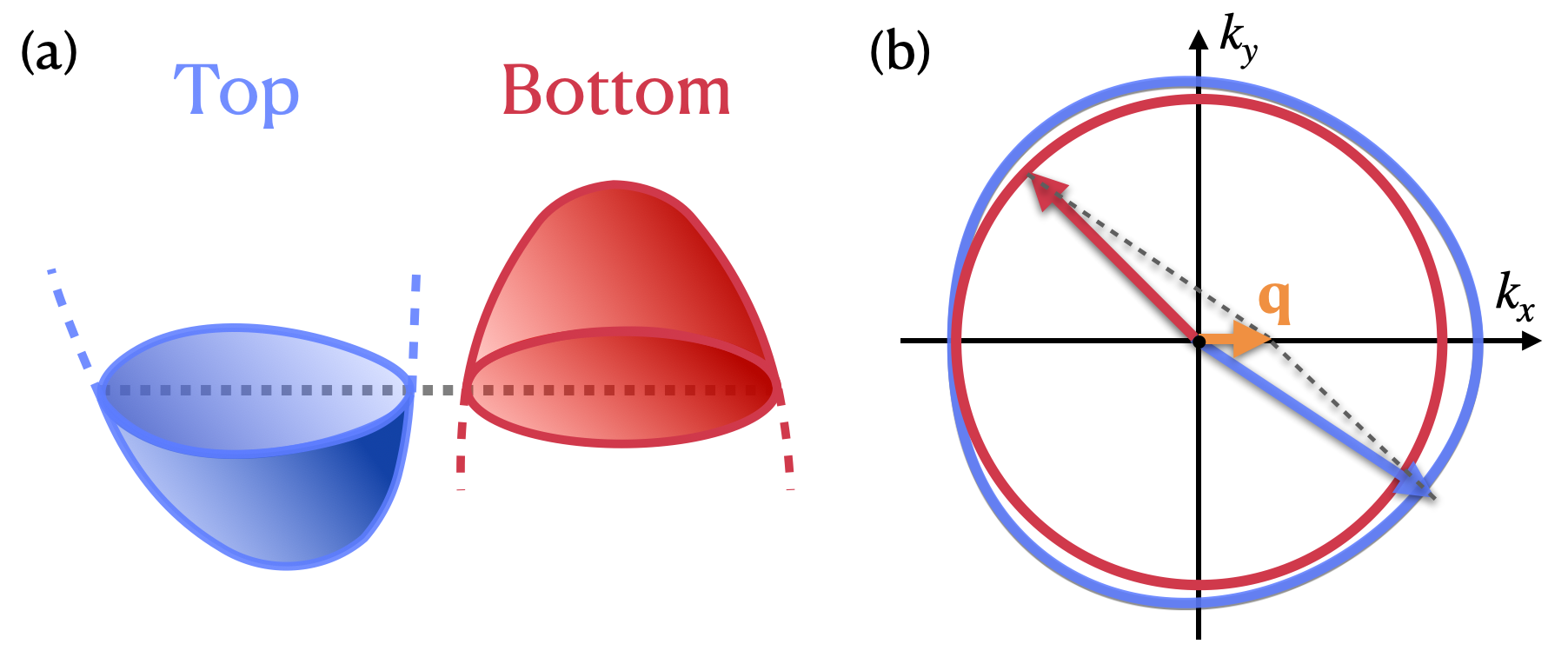}
\caption{\label{fig:setupsketch} \justifying (a) Electron dispersions and Fermi energies in both layers. Gating, spin–orbit coupling, and Zeeman fields create an electron Fermi sea in the top layer and a hole Fermi sea in the bottom layer. The gray dashed line marks the Fermi energy.\, (b) Schematic of electron and hole Fermi contours. The distortion and shift of the electron contour indicate finite-momentum $\bf q$ electron–hole pairing .}
\end{figure}

\isec{The model.} We consider a bilayer system composed of two-dimensional semiconductors whose conduction and valence bands have different SOC strengths. By electrostatic gating, the top layer is electron-doped into its conduction band, while the bottom layer is hole-doped into its valence band. The low-energy bands closest to the Fermi level in both layers, together with their corresponding dispersions, are illustrated in \figref{fig:setupsketch}{(a)}. To access the spin degree of freedom, we include both out-of-plane Zeeman field $J_{t,b}$ for both layers and an in-plane Zeeman field $B$. Combined with the Rashba SOC strengths
$\lambda_{e,h}$ in the top/bottom layer, the resulting effective noninteracting Hamiltonian at each layer can be written as:
\be
 H_a = (\frac{k^2}{2m_a}-\mu_a) + \lambda_{a}(k_x\sigma_y-k_y\sigma_x) + B\sigma_y + J_a\sigma_z \label{eq:ehfullH} 
\ee
Here, $a=e,h$ represents top electron-doped layer and bottom hole-doped layer respectively. We consider the case where $m_e=m_h$, and $\lambda_e>\lambda_h=0$ \footnote{Note when $\lambda_e=-\lambda_h$ and $\mu_e=\mu_h$, electron Fermi circle and hole Fermi circle are mirror image to each other. Under such condition, both finite-momentum pairing and supercurrent diode effect vanish.}. 
The out-of-plane Zeeman term splits the electron bands and hole bands, and one can assume low-energy carriers are spin polarized in the opposite direction. The dispersion relations at each layer are (We adopt natural units with $\hbar = c = 1$ throughout the paper and restore them when needed.)
\be
\xi_{a,k} = \frac{k^2}{2m_a}-\mu_a - |\lambda_{a}|\sqrt{(k_x+\frac{B}{\lambda_a})^2+k_y^2+(\frac{J_a}{\lambda_a})^2} \label{eq:ehdispersions}
\ee
Because of the non-vanishing  $\lambda_e$ and $B$ terms in the electron dispersion, both inversion ($\mathcal{P}$) and time-reversal ($\mathcal{T}$) symmetries are broken. Consequently, the electron band acquires a non-parabolic form, and its Fermi contour becomes non-circular, as illustrated in \figref{fig:setupsketch}{(b)}. In comparison with the red circular Fermi contour of holes in the bottom layer, the electron Fermi contour exhibits a simultaneous shift and distortion, which together energetically favor an exciton with a finite center-of-mass momentum, $\tb{q}$ ( orange arrow in \figref{fig:setupsketch}{(b)}), which is determined variationally through energy minimization (see later).
\newpara Interlayer Coulomb attraction between electrons in top layer and holes in bottom layer is essential for the formation of the exciton. The interlayer Coulomb interaction is $V(x,y)=-e^2/\epsilon\sqrt{x^2+y^2+d^2}$, with its Fourier transformation $V(p_x,p_y)=[2\pi e^2/\epsilon|\tb{p}|]e^{-|\tb{p}|d}$, where $\epsilon$ is taken as the dielectric constant of the intermediate insulating layer~\cite{ShaoDai2024Xbreakdown}, and $d$ is the distance between the layers in \z direction . Correspondingly, the second quantized interaction Hamiltonian reads:
\be
V_{eh} = -\frac{1}{\mathcal{V}}\sum_{\tb{k}_1,\tb{k}_2,\tb{p}} V(\tb{p})\psi^{\dagger}_{\tb{k}_1+\tb{p}} \chi^{\dagger}_{\tb{k}_2-\tb{p}} \chi_{\tb{k}_2} \psi_{\tb{k}_1} \label{Veh}
\ee
where $\psi,\chi$ are annihilation operators of electrons and holes, respectively. $\mathcal{V}$ is the volume of the sample. It is convenient to rewrite the interaction Hamiltonian in the electron--hole pairing channel and expand the interaction matrix element in angular-momentum channels using the Fourier basis in two-dimensional polar coordinates~\cite{Leggett2006Qliquid,Shankar1994RGfermion}, i.e.,
\[
V_{eh} = -\frac{1}{\mathcal{V}}\sum_{\tb{k},\tb{k}^{\prime},\tb{q}} V(\tb{k}-\tb{k}^{\prime})\psi^{\dagger}_{\tb{k}+\tb{q}/2} \chi^{\dagger}_{-\tb{k}+\tb{q}/2} \chi_{-\tb{k}^{\prime}+\tb{q}/2}\psi_{\tb{k}^{\prime}+\tb{q}/2} 
\]
where 
$V(\tb{k}-\tb{k}^{\prime})=\sum_{m=0,\pm 1,\pm2,...}V_m(|\tb{k}|,|\tb{k}^{\prime}|)\e^{im(\phi-\phi^{\prime})}$, and $\phi,\phi^{\prime}$ are azimuthal angles of $\tb{k},\tb{k}^{\prime}$, respectively \footnote{In the BCS regime of an exciton condensate, the interlayer Coulomb interaction is further screened by plasmonic modes of the carriers in the two layers. Once such screening effects are included, the screened interaction no longer has an infrared divergence at small momentum; see Ref.~\cite{LozovikYudson1976ehSC}.}.
Following the standard pairing approximation~\cite{SchriefferSCtextbook}, we retain a single center-of-mass momentum $\bq$ and keep only the dominant $m=0$ ($s$-wave) pairing channel~\cite{LozovikYudson1976ehSC,Shevchenko1976ehSC,FilShevchenko2018ehSCreview}. The effective interaction Hamiltonian then reduces to
\be
V_{pair} = -\frac{g}{\mathcal{V}}\sum_{k,k^\prime} \psi^{\dagger}_{\tb{k}+\tb{q}/2} \chi^{\dagger}_{-\tb{k}+\tb{q}/2} \chi_{-\tb{k}^\prime+\tb{q}/2} \psi_{\tb{k}^\prime+\tb{q}/2} \label{eq:Vpairing}
\ee
where $g = V_{m=0}({|\tb{k}|=|\tb{k}^\prime|=k_F})$ is the effective $s$-wave interaction strength, which is suppressed as the interlayer separation $d$ increases; here $k_F=(4\pi)^{-1}\sum_{a=e,h}\int d\phi\,|\tb k|_{\xi_{a,\tb{k}}=0}$ denotes the average Fermi momentum for an anisotropic Fermi surface.
As discussed in depth by Lozovik and Yudson~\cite{LozovikYudson1976ehSC}, the effective coupling $g$ does not generally admit a closed-form expression, although analytical results can be obtained in certain limiting regimes. In the numerical calculations presented below, we therefore adopt a phenomenological value of $g$ inferred from experimentally relevant superfluid transition temperatures reported for TMD-based exciton condensates.

\isec{Ginzburg-Landau theory and finite momentum condensate.} Based on the above, the second-quantized effective Hamiltonian is $H=H_0+V_{pair}$, where
\be
H_0 = \sum_{\tb{k}} \xi_{e,\tb{k}}\psi^{\dagger}_{\tb{k}}\psi_{\tb{k}} + \xi_{h,\tb{k}}\chi^{\dagger}_{\tb{k}}\chi_{\tb{k}} \label{eq:H0}
\ee
and $V_{pair}$ is given in \eqref{eq:Vpairing}. We take the Fulde-Ferrell-type~\cite{FF1964} ansatz solution for the excitonic order parameter field $\Delta(\tb{x})=\Delta\,e^{i\tb{q}\cdot\tb{x}}$, where \bq is the exciton momentum. Performing a Hubbard-Stratonovich transformation yields the effective Euclidean action $S[\bar{\Delta},\Delta]$:
\be
e^{-S[\bar{\Delta},\Delta]}=e^{-\int_{0}^{\beta}\,d\tau\frac{\mathcal{V}}{g}(\bar{\Delta}\Delta)}\int\,D(\bar{\psi},\psi)D(\bar{\chi},\chi)e^{-S_{BdG}}\label{eq:Sdeltaeff}
\ee
where $S_{BdG}=\int\,d\tau\sum_{\tb{k}}[\bar{\psi}_{\tb{k}}(\partial_\tau+\xi_{e,\tb{k}})\psi_{\tb{k}}+\bar{\chi}_{\tb{k}}(\partial_\tau+\xi_{h,\tb{k}})\chi_{\tb{k}}-(\bar{\Delta}\chi_{-\tb{k}+\tb{q}/2}\psi_{\tb{k}+\tb{q}/2}+h.c.)]$. Near the phase transition point (i.e., $\Delta\ll k_B T_c$), we keep the leading terms in $S[\bar{\Delta},\Delta]=T^{-1}F[\bar{\Delta},\Delta]$, and derive Ginzburg-Landau free energy \cite{Kleinert1978collectiveQF,Altland2010textbook,Laine2016TFT}
\be
F[\bar\Delta, \Delta] = \alpha(T,\textbf{q})|\Delta|^2 + \half \beta(T,\textbf{q})|\Delta|^4 + O(|\Delta|^6) \label{eq:FDeltaGL}
\ee
where 
\be
\alpha(T,\textbf{q}) &=& \frac{\mathcal{V}}{g}+\sum_{n,\tb{k}}\frac{T}{(-i\omega_n+\xi_{e,\tb{k}+\frac{\tb{q}}{2}})(-i\omega_n-\xi_{h,-\tb{k}+\frac{\tb{q}}{2}})} \nonumber \\
\beta(T,\textbf{q}) &=& \sum_{n,\tb{k}}\frac{T}{(-i\omega_n+\xi_{e,\tb{k}+\frac{\tb{q}}{2}})^2(-i\omega_n-\xi_{h,-\tb{k}+\frac{\tb{q}}{2}})^2} \label{eq:GLbeta}
\ee
and $\omega_n=(2n+1)\pi T$ with $n=0,\pm1,\pm2,...$ are the fermionic Matsubara frequencies. It is evident from \eqref{eq:GLbeta} that, if $\xi_{e,\tb{k}}=\xi_{e,-\tb{k}}$ and $\xi_{h,\tb{k}}=\xi_{h,-\tb{k}}$, both $\alpha$ and $\beta$ are even functions of \bq, and therefore no odd-exponent terms appear in the small-$q$ expansion. As previously pointed out, the presence of terms odd in $\mathbf{q}$ is essential for realizing finite-momentum condensates and supercurrent diode effect~\cite{YuanFu2022SDE,HeNagaosa2022SDE,Daido2022SDE}. Therefore, our model requires $\xi_{e,\tb{k}}\neq\xi_{e,\tb{-k}}$, which occurs when both inversion and time-reversal symmetries are broken induced by a nonzero in-plane Zeeman field $B$.
\begin{figure}[htbp]
\includegraphics[width=0.46\textwidth]{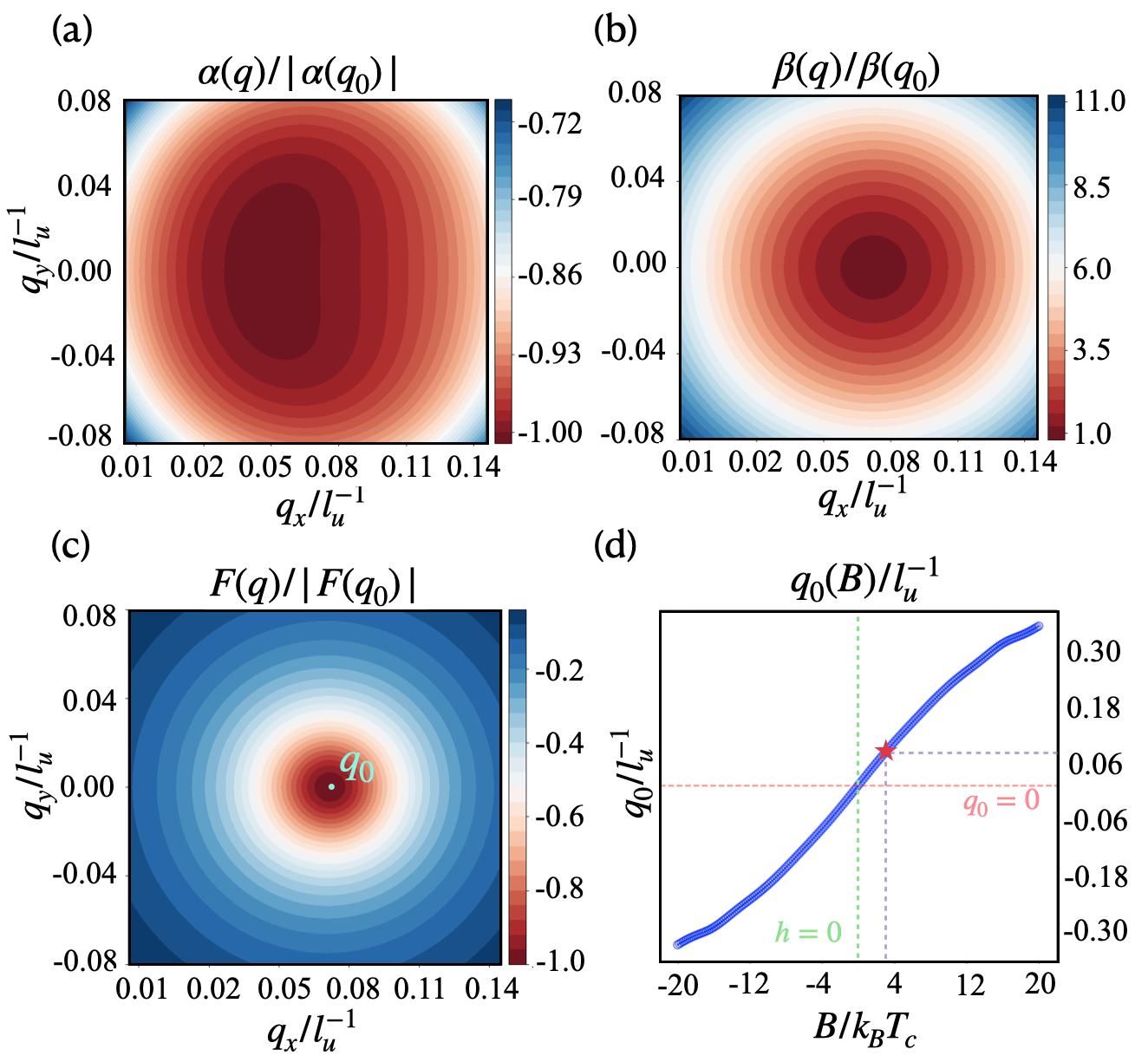}
\caption{\label{fig:bcslandscape} \justifying (a) Exciton condensate momentum $q_0$ as a function of $B/k_BT_c$ at $0.90T_c$. At $B/k_BT_c$=3.0, we calculate the Ginzburg-Landau coefficients of the condensate as a function of exciton momentum $(q_x,q_y)$, based on which we calculate free energy and extract $q_0$. \,(b, c, d) Normalized values of $\alpha(\tb{q}), \beta(\tb{q})$ and $F(\tb{q})$ at $T/T_c=0.90, B/k_BT_c =3.0$, where red dot represents exciton momentum $\tb{q}_0$ that minimizes $F(\tb{q})$. $E_u=k_BT_c$ and $l_u=\hbar/\sqrt{m_ek_BT_c}$ are energy and length units, respectively. }
\end{figure}
\newpara In the Ginzburg-Landau framework, physical quantities of interest can be expressed in terms of $\alpha$ and $\beta$. For instance, free energy $F(\tb{q})=-[\alpha(\tb{q})]^2/\beta(\tb{q})$, and the supercurrent density $J_i(\tb{q})=(\hbar\mathcal{V})^{-1}\partial F(\tb{q})/\partial q_i$. Based on expressions \eqref{eq:GLbeta}, we numerically calculated the $q_x$ and $q_y$-dependence  of $\alpha(T,\tb{q})$ and $\beta(T,\tb{q})$ at $T/T_c=0.90$, $B/k_BT_c=3.0$ (\figref{fig:bcslandscape}{(a, b)}). Within the exciton momentum range shown in the figure, $\alpha<0$ and $\beta>0$, ensures condensation. For exciton momenta away from the center of the figure, excitons flow, and the order parameter $\Delta=|\alpha|/\beta$ is suppressed as expected. Furthermore, we plot the $q_x,q_y$-dependence of the free energy $F$ in \figref{fig:bcslandscape}{(c)}. Treating $\tb{q}=(q_x,q_y)$ as a variational parameter, the exciton condensate at the momentum $\tb{q}_0=(q_0,0)$ at which the free energy is minimized. We show the $B$-dependence of the optimal $q_0$ in \figref{fig:bcslandscape}{(d)}, showing a monotonically increase with the in-plane Zeeman field $B$.

\begin{figure}[htbp]
\includegraphics[width=0.46\textwidth]{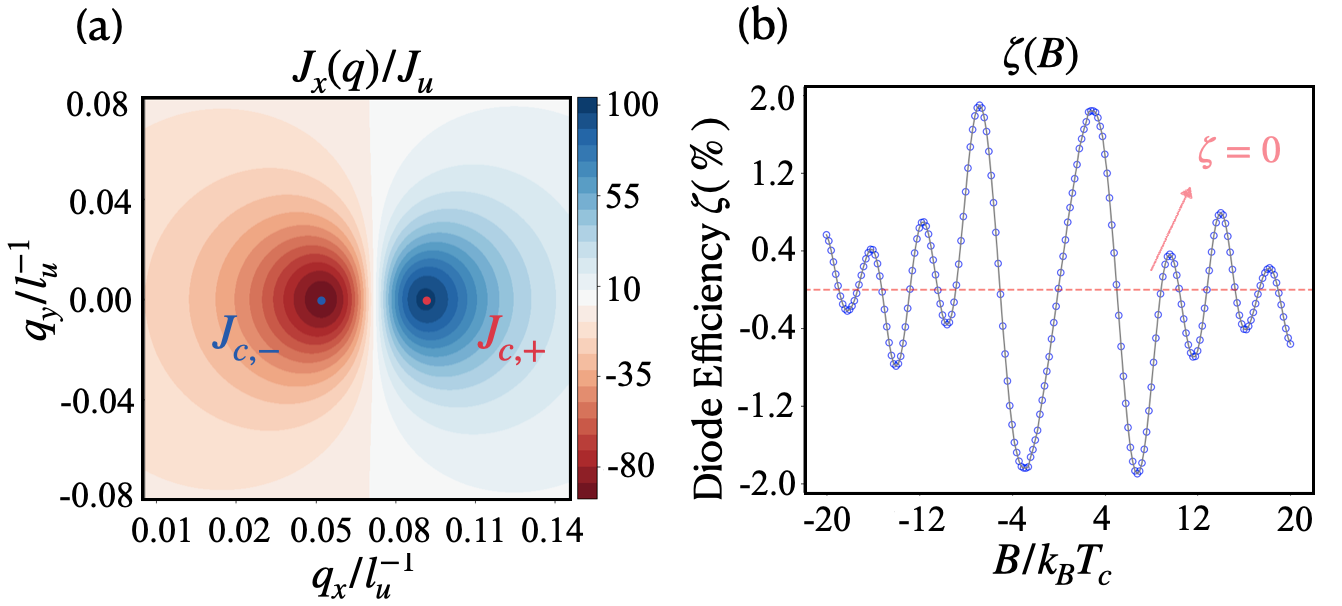}
\caption{\label{fig:diodeplot} \justifying (a) Supercurrent density $J_x$ as a function of exciton momentum $q_x$ at $T/T_c=0.90, B/k_BT_c =3.0$.  \, 
(b) At $T/T_c=0.90$, diode efficiency $\eta$ as a function of $B/k_BT_c$.}
\end{figure}

\isec{Coherent Exciton diode effect.} The excitonic supercurrent density as a function of exciton momentum $\tb{q}$ is $J_i(\tb{q})=(\hbar\mathcal{V})^{-1}\partial F(\tb{q})/\partial q_i$, which vanishes in ground state at $\tb{q}=\tb{q}_0$. At $0.9\,T_c$, we focus on the \x-direction, where the mirror symmetry is broken, and plot the numerical values $J_x(q_x,q_y)$ at $B/k_BT_c=3.0$ in \figref{fig:diodeplot}{(a)}. 
As $q_x$ is varied to generate a nonzero supercurrent $J_x$, we obtain a minimum $J_{c-}$ and a maximum $J_{c+}$, which we identify as critical currents in two opposite directions along the \x axis~\cite{Bardeen1962critical,TinkhamSCtextbook,deGennesSCtextbook}.
In the presence of either $\mathcal{P}$ or $\mathcal{T}$ symmetry, the relation $|J_{c+}| = |J_{c-}|$ must hold. However, when both symmetries are broken—as in our model—this constraint no longer applies, and we indeed find $J_{c+} \neq J_{c-}$. Each exciton possesses an electric dipole oriented along the $-z$ direction, so the excitonic diode effect in the bilayer may be naturally viewed as a supercurrent diode effect of electric dipoles~\cite{Jiang2015ediopolesc,shao2025electromagnetic}.
\newpara In \figref{fig:diodeplot}{(a)}, we extract $J_{c+}/J_u=98.4$, $J_{c-}/J_u=-94.8$. 
Using typical model parameters, we estimate $|J_{c+}| - |J_{c-}|$ to be on the order of $10\,(\mu\mathrm{A})(\mu\mathrm{m})^{-1}$, which is well within experimental reach. The diode efficiency, defined as $\eta = (|J_{c+}| - |J_{c-}|)/(|J_{c+}| + |J_{c-}|)$, is shown as a function of the external in-plane field $h$ in \figref{fig:diodeplot}{(b)}. Interestingly, $\eta$ does not increase monotonically with $h$ but instead exhibits irregular oscillations. At certain values of $h$, $\eta$ even vanishes, showing the absence of a supercurrent diode effect. This behavior demonstrates that, while breaking both $\mathcal{P}$ and $\mathcal{T}$ symmetries is necessary for the emergence of a supercurrent diode effect, it is not sufficient. 
Our numeric simulations reveal a nonlinear oscillatory behavior in the diode efficiency, displaying quasiperiodicity over scales of several $k_B T_c$. This nonlinear behavior stems from our investigation of the condensate's breakdown current~\cite{supplemental_material}. 


\isec{Frictional Coulomb drag in the normal phase.} When the excitonic supercurrent exceeds the critical current, the drag ratio does not drop to 0 but remains a finite small value, as shown in \figref{fig:dragsketch}{(b)}. Such remaining drag effect away from the condensate phase is called frictional Coulomb drag~\cite{Narozhny2016dragreview}. This originates from momentum transfer between the charge carriers in different layers, which has been studied in a series delicate works using kinetic theory~\cite{Jauho1993Coulombdrag,Rojo1999Electrondrag,Narozhny2016dragreview}. Flowing charge carriers in the drive layer (layer 1) will exert a force on carriers in the drag layer (layer 2), which in the steady state has to be balanced out by either electrical force or frictional force. 

In our system, the top layer contains electrons with a dispersion that breaks both 
$\mathcal{P}$ and $\mathcal{T}$ symmetries, while the bottom layer consists of holes with parabolic dispersion, as shown in \eqref{eq:ehdispersions}. We calculate the drag ratio in the x-direction. To the linear order, the current in the top layer $j_1$ and bottom layer $j_2$ satisfied $j_2=\zeta\cdot j_1$, where the expression for $\zeta$ is derived in the Supplemental Materials~\cite{supplemental_material} 

\be
\zeta= -\frac{n_h\bar{m}_{e,x}}{n_em_h}\cdot\frac{1}{(\tau_D^{\Rmnum{1}}/\tau_D^{\Rmnum{2}})+(\tau_D^{\Rmnum{1}}/\tau_2)} \label{eq:dragzeta}
\ee
    where $n_{e,h}$ are charge carrier densities in either layer. $\bar{m}_{e,x}^{-1}=n_e^{-1}\int\frac{d^2k}{(2\pi)^2}f_0(\xi_{e,\tb{k}})\partial_{k_x}^2\xi_{e,\tb{k}}$ is the ensemble average of inverse effective mass in \x direction with $f_0(\xi)=(\e^{\beta\xi}+1)^{-1}$ the Fermi-Dirac distribution function. The scattering time $\tau_{D}^{\Rmnum{1},\Rmnum{2}}$ are provided in Supplemental materials~\cite{supplemental_material}. For our parameters, a rough estimate yields $\tau_{D}^{\Rmnum{1}} \approx \tau_{D}^{\Rmnum{2}} \approx 1\times10^{-7}\,\mathrm{s}$, assuming the Fermi energies are $E_{Fe} \approx E_{Fh} \approx 10\mathrm{meV}$ at $T = 10\,\mathrm{K}$. In a clean sample, one often takes $\tau_2 \approx 1\text{–}2\times10^{-9}\,\mathrm{s}$, leading to a drag ratio of $\zeta \approx 1\%$. As the Fermi energies decrease, both $\tau_D^{-1}$ and $\zeta$ increase, consistent with Ref.~\cite{Jauho1993Coulombdrag}. It is important to note that this analysis can be extended to higher-order processes, where simultaneous breaking of $\mathcal{P}$ and $\mathcal{T}$ symmetries may lead to a nonlinear, nonreciprocal drag effect~\cite{Rikken2001electricalMCA,Ideue2017bulkpolarMCA}. While these higher-order effects are beyond the main focus of this work, a brief discussion is provided in the Supplemental Materials~\cite{supplemental_material}.

\isec{Conclusion \& Outlook.}
We proposed the coherent-exciton diode effect, which gives a robust signature of interlayer exciton condensation through a nonreciprocal Coulomb drag response. 
Using typical experimental parameters, i.e., critical temperature $T_c = 10\,\mathrm{K}$ and effective electron mass of $m_e \approx 0.2\,m_0$ ($m_0$ is the bare electron mass), the characteristic exciton condensate momentum is estimated as 
$q_0 = 0.2\,l_u^{-1} \approx 0.2\sqrt{m_e k_B T_c / \hbar^2}$, corresponding to a phase modulation wavelength of $\lambda \approx 600\,\mathrm{nm}$—well within current experimental reach. 
For a sample with a width of $1\,\mu\mathrm{m}$, the critical current can reach $I_c \approx 1\,\mathrm{mA}$, resulting in an expected current difference of $\delta I \approx 10\,\mu\mathrm{A}$ given a diode efficiency of $\eta = 1\%$. 
Such signals should be readily detectable in two-dimensional layered materials with strong spin–orbit coupling, such as transition-metal dichalcogenides (TMD) monolayers~\cite{Mak2010MoS2,XiaoYao2012TMDvalley,ReganMacDengWang2022XTMDreview}. In particular, to realize Rashba spin-orbit coupling instead of Ising spin-orbit coupling only, one can use Janus TMD monolayer, which breaks inversion symmetry in \z direction intrinsically and introduce a Rashba SOC term in the electronic Hamiltonian~\cite{Lu2017janus,Tang2022Janus}. In the supplemental material, we elaborated on material realization and propose a concrete experimental setup~\cite{supplemental_material}.

The electron–hole bilayer system, with its remarkable tunability, provides an ideal platform for realizing a broad spectrum of exotic quantum states. For instance, excitons have been observed in both the weak-coupling BCS limit and the strong-coupling BEC limit~\cite{LiuDean2022crossover}, unlocking new possibilities for exploring the BCS–BEC crossover in solid-state systems~\cite{Chen2005bcsbec}. An in-plane electric field can dissociate exciton pairs, inducing a many-body breakdown mechanism~\cite{ShaoDai2024Xbreakdown}, and several other experimental signatures have also been proposed and detected~\cite{ShaoDai2024quantumoscillationexciton, HanMak2025quantumoscillationexciton, QiWang2025competitionexcitonQSH, ShaoDai2025EMresponseexciton}.
Moreover, gating imbalance in electron–hole bilayers could lead to a wealth of intriguing physics, such as excitonic FFLO states~\cite{VarleyLee2016imbalancedeh, Seradjeh2012imbalancedeh,KumarSenthil2024XDW}, excitonic density waves~\cite{WuWang2026XDWdiode}, or time-reversal broken topologically ordered ground states~\cite{WangDu2023XTO}. When electron–hole densities are imbalanced yet commensurate, the system favors the formation of exotic three-body bound states, such as trions~\cite{OuyangShi2025trionnews, QiWang2025trion, NguyenShanMak2025trion, DaiFu2024trion, ZerbaKnap2024mixturepwave}. 
In this rich landscape of physics, our proposed coherent-exciton diode effect opens a novel and intriguing avenue for detecting and manipulating exotic excitonic states, offering a promising route to realize new quantum devices based on excitonic transport and nonreciprocal behaviors.

\isec{Acknowledgment.} The authors are grateful to Gabriel Cardoso for useful discussions throughout the project and detailed suggestions to the manuscript. The authors acknowledge Thors Hans Hansson for insightful discussions and constructive suggestions. J.-X. Hui thanks Xiaodong He and Xiaotong Chen for their generous help with numerical calculations. The authors also appreciate James Jun He, Noah F.Q. Yuan, Liguo Ma, Xiaoxue Liu and Shengwei Jiang for useful discussions. This work was supported by National Natural Science Foundation of China (NSFC) under Grant No. 12374332, Quantum Science and Technology-National Science and Technology Major Project No. 2021ZD0301900, Cultivation Project of Shanghai Research Center for Quantum Sciences Grant No.LZPY2024, and Shanghai Science and Technology Innovation Action Plan Grant No. 24LZ1400800.

\isec{Data availability.} The data that support the findings of this article are publicly available\cite{DVN/NYXL70_2026}.

\bibliography{XSDE.bib}

@article{LozovikYudson1976ehSC,
  title={A new mechanism for superconductivity: pairing between spatially separated electrons and holes},
  author={Lozovik, Yu E and Yudson, VI},
  journal={Zh. Eksp. Teor. Fiz},
  volume={71},
  pages={738},
  year={1976}
}

@article{Shevchenko1976ehSC,
  title={Theory of superconductivity of systems with pairing of spatially separated electrons and holes},
  author={Shevchenko, SI},
  journal={Soviet Journal of Low Temperature Physics},
  volume={2},
  number={4},
  pages={251--257},
  year={1976},
  publisher={American Institute of Physics}
}

@article{FilShevchenko2018ehSCreview,
  title={Electron-hole superconductivity},
  author={Fil, DV and Shevchenko, SI},
  journal={Low Temperature Physics},
  volume={44},
  number={9},
  pages={867--909},
  year={2018},
  publisher={AIP Publishing}
}

@article{Nandi2012XCPCD,
  title={Exciton condensation and perfect Coulomb drag},
  author={Nandi, D and Finck, ADK and Eisenstein, JP and Pfeiffer, LN and West, KW},
  journal={Nature},
  volume={488},
  number={7412},
  pages={481--484},
  year={2012},
  publisher={Nature Publishing Group UK London}
}

@article{NguyenShanMak2025XCPCD,
  title={Perfect Coulomb drag in a dipolar excitonic insulator},
  author={Nguyen, Phuong X and Ma, Liguo and Chaturvedi, Raghav and Watanabe, Kenji and Taniguchi, Takashi and Shan, Jie and Mak, Kin Fai},
  journal={Science},
  volume={388},
  number={6744},
  pages={274--278},
  year={2025},
  publisher={American Association for the Advancement of Science}
}

@article{Jiang2015ediopolesc,
  title={Theory for electric dipole superconductivity with an application for bilayer excitons},
  author={Jiang, Qing-Dong and Bao, Zhi-qiang and Sun, Qing-Feng and Xie, XC},
  journal={Scientific reports},
  volume={5},
  number={1},
  pages={11925},
  year={2015},
  publisher={Nature Publishing Group UK London}
}

@article{Frenkel1931,
  title={On the transformation of light into heat in solids. I},
  author={Frenkel, Jacov},
  journal={Physical Review},
  volume={37},
  number={1},
  pages={17},
  year={1931},
  publisher={APS}
}

@article{Sturge1962optical,
  title={Optical absorption of gallium arsenide between 0.6 and 2.75 eV},
  author={Sturge, MoDo},
  journal={Physical Review},
  volume={127},
  number={3},
  pages={768},
  year={1962},
  publisher={APS}
}

@article{Hopfield1958excitonpolariton,
  title={Theory of the contribution of excitons to the complex dielectric constant of crystals},
  author={Hopfield, JJ},
  journal={Physical Review},
  volume={112},
  number={5},
  pages={1555},
  year={1958},
  publisher={APS}
}

@article{Snoke2002excitonpolariton,
  title={Spontaneous Bose coherence of excitons and polaritons},
  author={Snoke, David},
  journal={Science},
  volume={298},
  number={5597},
  pages={1368--1372},
  year={2002},
  publisher={American Association for the Advancement of Science}
}

@article{Blatt1962exciton,
  title={Bose-Einstein condensation of excitons},
  author={Blatt, John M and B{\"o}er, KW and Brandt, Werner},
  journal={Physical Review},
  volume={126},
  number={5},
  pages={1691},
  year={1962},
  publisher={APS}
}

@article{Keldysh1964exciton,
  title={Possible instability of the semimetallic state toward coulomb interaction},
  author={Keldysh, LV and Kopaev, You V},
  journal={Fiz. Tverd. Tela},
  volume={6},
  pages={2791},
  year={1964}
}

@article{Jerome1967EI,
  title={Excitonic insulator},
  author={Jerome, Denis and Rice, TM and Kohn, W},
  journal={Physical Review},
  volume={158},
  number={2},
  pages={462},
  year={1967},
  publisher={APS}
}

@article{Pereira2022topoexciton,
  title={Topological excitons},
  author={Pereira, Vitor M},
  journal={Nature Physics},
  volume={18},
  number={1},
  pages={6--7},
  year={2022},
  publisher={Nature Publishing Group UK London}
}

@article{WuMacdonald2015theory,
  title={Theory of two-dimensional spatially indirect equilibrium exciton condensates},
  author={Wu, Feng-Cheng and Xue, Fei and MacDonald, AH},
  journal={Physical Review B},
  volume={92},
  number={16},
  pages={165121},
  year={2015},
  publisher={APS}
}

@article{Moon2025VdWexcitonreview,
  title={Exciton condensate in van der Waals layered materials},
  author={Moon, Byoung Hee and Mondal, Ashok and Efimkin, Dmitry K and Lee, Young Hee},
  journal={Nature Reviews Physics},
  pages={1--14},
  year={2025},
  publisher={Nature Publishing Group UK London}
}

@article{Eisenstein2014excitonQHreview,
  title={Exciton condensation in bilayer quantum Hall systems},
  author={Eisenstein, JP},
  journal={Annu. Rev. Condens. Matter Phys.},
  volume={5},
  number={1},
  pages={159--181},
  year={2014},
  publisher={Annual Reviews}
}

@article{YangDaiLi2025EIreview,
  title={Exciton Insulators in Two-dimensional Systems},
  author={Yang, Huaiyuan and Dai, Xi and Li, Xin-Zheng},
  journal={Chinese Physics B},
  year={2025}
}

@article{ZhuDasSarma2024bilayer,
  title={Interaction and coherence in two-dimensional bilayers},
  author={Zhu, Jihang and Das Sarma, Sankar},
  journal={Physical Review B},
  volume={109},
  number={8},
  pages={085129},
  year={2024},
  publisher={APS}
}

@article{NguyenMak2025quantumoscillationEI,
  title={Quantum oscillations in a dipolar excitonic insulator},
  author={Nguyen, Phuong X and Chaturvedi, Raghav and Zou, Bo and Watanabe, Kenji and Taniguchi, Takashi and MacDonald, Allan H and Mak, Kin Fai and Shan, Jie},
  journal={Nature Materials},
  pages={1--7},
  year={2025},
  publisher={Nature Publishing Group UK London}
}

@article{Karni2022trARPESexciton,
  title={Structure of the moir{\'e} exciton captured by imaging its electron and hole},
  author={Karni, Ouri and Barr{\'e}, Elyse and Pareek, Vivek and Georgaras, Johnathan D and Man, Michael KL and Sahoo, Chakradhar and Bacon, David R and Zhu, Xing and Ribeiro, Henrique B and O’Beirne, Aidan L and others},
  journal={Nature},
  volume={603},
  number={7900},
  pages={247--252},
  year={2022},
  publisher={Nature Publishing Group UK London}
}

@article{ShaoDai2024quantumoscillationexciton,
  title={Quantum oscillations in an excitonic insulating electron-hole bilayer},
  author={Shao, Yuelin and Dai, Xi},
  journal={Physical Review B},
  volume={109},
  number={15},
  pages={155107},
  year={2024},
  publisher={APS}
}

@article{HanMak2025quantumoscillationexciton,
  title={Quantum oscillations between excitonic and quantum spin Hall insulators in moir$\backslash$'e WSe2},
  author={Han, Zhongdong and Xia, Yiyu and Watanabe, Kenji and Taniguchi, Takashi and Mak, Kin Fai and Shan, Jie},
  journal={arXiv preprint arXiv:2509.19287},
  year={2025}
}

@article{QiWang2025competitionexcitonQSH,
  title={Competition between excitonic insulators and quantum Hall states in correlated electron--hole bilayers},
  author={Qi, Ruishi and Li, Qize and Zhang, Zuocheng and Nie, Jiahui and Zou, Bo and Cui, Zhiyuan and Kim, Haleem and Sanborn, Collin and Chen, Sudi and Xie, Jingxu and others},
  journal={Nature Materials},
  pages={1--7},
  year={2025},
  publisher={Nature Publishing Group UK London}
}

@article{ShaoDai2025EMresponseexciton,
  title={Electromagnetic responses of bilayer excitonic insulators},
  author={Shao, Yuelin and Shi, Hao and Dai, Xi},
  journal={arXiv preprint arXiv:2509.02142},
  year={2025}
}

@article{MaMak2021EIindoublelayer,
  title={Strongly correlated excitonic insulator in atomic double layers},
  author={Ma, Liguo and Nguyen, Phuong X and Wang, Zefang and Zeng, Yongxin and Watanabe, Kenji and Taniguchi, Takashi and MacDonald, Allan H and Mak, Kin Fai and Shan, Jie},
  journal={Nature},
  volume={598},
  number={7882},
  pages={585--589},
  year={2021},
  publisher={Nature Publishing Group UK London}
}

@article{WangMak2019evidenceXC,
  title={Evidence of high-temperature exciton condensation in two-dimensional atomic double layers},
  author={Wang, Zefang and Rhodes, Daniel A and Watanabe, Kenji and Taniguchi, Takashi and Hone, James C and Shan, Jie and Mak, Kin Fai},
  journal={Nature},
  volume={574},
  number={7776},
  pages={76--80},
  year={2019},
  publisher={Nature Publishing Group UK London}
}

@article{Fogler2014BX,
  title={High-temperature superfluidity with indirect excitons in van der Waals heterostructures},
  author={Fogler, MM and Butov, LV and Novoselov, KS},
  journal={Nature communications},
  volume={5},
  number={1},
  pages={4555},
  year={2014},
  publisher={Nature Publishing Group UK London}
}

@article{QiWang2025PCD,
  title={Perfect Coulomb drag and exciton transport in an excitonic insulator},
  author={Qi, Ruishi and Joe, Andrew Y and Zhang, Zuocheng and Xie, Jingxu and Feng, Qixin and Lu, Zheyu and Wang, Ziyu and Taniguchi, Takashi and Watanabe, Kenji and Tongay, Sefaattin and others},
  journal={Science},
  volume={388},
  number={6744},
  pages={278--283},
  year={2025},
  publisher={American Association for the Advancement of Science}
}

@article{ReganMacDengWang2022XTMDreview,
  title={Emerging exciton physics in transition metal dichalcogenide heterobilayers},
  author={Regan, Emma C and Wang, Danqing and Paik, Eunice Y and Zeng, Yongxin and Zhang, Long and Zhu, Jihang and MacDonald, Allan H and Deng, Hui and Wang, Feng},
  journal={Nature Reviews Materials},
  volume={7},
  number={10},
  pages={778--795},
  year={2022},
  publisher={Nature Publishing Group UK London}
}

@article{XieMac2018Xresevoir,
  title={Electrical reservoirs for bilayer excitons},
  author={Xie, Ming and MacDonald, Allan H},
  journal={Physical review letters},
  volume={121},
  number={6},
  pages={067702},
  year={2018},
  publisher={APS}
}

@article{QiMacWang20262compXC,
  title={Observation of two-component exciton condensates in an excitonic insulator},
  author={Qi, Ruishi and Li, Qize and Nie, Jiahui and Xia, Ruichen and Kim, Haleem and Lim, Hyungbin and Xie, Jingxu and Taniguchi, Takashi and Watanabe, Kenji and Crommie, Michael F and others},
  journal={arXiv preprint arXiv:2603.15443},
  year={2026}
}

@article{WuWang2026XDWdiode,
  title={Spontaneously formed excitonic density wave with vortex-antivortex lattice in twisted semiconductor bilayers},
  author={Wu, Deguang and Xue, Yiran and Wang, Baigeng and Wang, Rui and Xing, DY},
  journal={arXiv preprint arXiv:2603.28164},
  year={2026}
}

@article{Ando2020SDEobservation,
  title={Observation of superconducting diode effect},
  author={Ando, Fuyuki and Miyasaka, Yuta and Li, Tian and Ishizuka, Jun and Arakawa, Tomonori and Shiota, Yoichi and Moriyama, Takahiro and Yanase, Youichi and Ono, Teruo},
  journal={Nature},
  volume={584},
  number={7821},
  pages={373--376},
  year={2020},
  publisher={Nature Publishing Group UK London}
}

@article{Nadeem2023SDEreview,
  title={The superconducting diode effect},
  author={Nadeem, Muhammad and Fuhrer, Michael S and Wang, Xiaolin},
  journal={Nature Reviews Physics},
  volume={5},
  number={10},
  pages={558--577},
  year={2023},
  publisher={Nature Publishing Group UK London}
}

@article{YuanFu2022SDE,
  title={Supercurrent diode effect and finite-momentum superconductors},
  author={Yuan, Noah FQ and Fu, Liang},
  journal={Proceedings of the National Academy of Sciences},
  volume={119},
  number={15},
  pages={e2119548119},
  year={2022},
  publisher={National Academy of Sciences}
}

@article{HeNagaosa2022SDE,
  title={A phenomenological theory of superconductor diodes},
  author={He, James Jun and Tanaka, Yukio and Nagaosa, Naoto},
  journal={New Journal of Physics},
  volume={24},
  number={5},
  pages={053014},
  year={2022},
  publisher={IOP Publishing}
}

@article{Daido2022SDE,
  title={Intrinsic superconducting diode effect},
  author={Daido, Akito and Ikeda, Yuhei and Yanase, Youichi},
  journal={Physical Review Letters},
  volume={128},
  number={3},
  pages={037001},
  year={2022},
  publisher={APS}
}

@article{JiangHu2022SDE,
  title={Superconducting diode effects},
  author={Jiang, Kun and Hu, Jiangping},
  journal={Nature Physics},
  volume={18},
  number={10},
  pages={1145--1146},
  year={2022},
  publisher={Nature Publishing Group UK London}
}

@article{ZhangHuJiang2022JDE,
  title={General theory of Josephson diodes},
  author={Zhang, Yi and Gu, Yuhao and Li, Pengfei and Hu, Jiangping and Jiang, Kun},
  journal={Physical Review X},
  volume={12},
  number={4},
  pages={041013},
  year={2022},
  publisher={APS}
}

@article{LeLinWu2024KagomeSDE,
  title={Superconducting diode effect and interference patterns in kagome CsV3Sb5},
  author={Le, Tian and Pan, Zhiming and Xu, Zhuokai and Liu, Jinjin and Wang, Jialu and Lou, Zhefeng and Yang, Xiaohui and Wang, Zhiwei and Yao, Yugui and Wu, Congjun and others},
  journal={Nature},
  volume={630},
  number={8015},
  pages={64--69},
  year={2024},
  publisher={Nature Publishing Group UK London}
}

@article{Lin2022zeroBSDE,
  title={Zero-field superconducting diode effect in small-twist-angle trilayer graphene},
  author={Lin, Jiang-Xiazi and Siriviboon, Phum and Scammell, Harley D and Liu, Song and Rhodes, Daniel and Watanabe, K and Taniguchi, T and Hone, James and Scheurer, Mathias S and Li, JIA},
  journal={Nature Physics},
  volume={18},
  number={10},
  pages={1221--1227},
  year={2022},
  publisher={Nature Publishing Group UK London}
}

@article{Wu2022zeroBJDE,
  title={The field-free Josephson diode in a van der Waals heterostructure},
  author={Wu, Heng and Wang, Yaojia and Xu, Yuanfeng and Sivakumar, Pranava K and Pasco, Chris and Filippozzi, Ulderico and Parkin, Stuart SP and Zeng, Yu-Jia and McQueen, Tyrel and Ali, Mazhar N},
  journal={Nature},
  volume={604},
  number={7907},
  pages={653--656},
  year={2022},
  publisher={Nature Publishing Group UK London}
}

@article{Trahms2023JDE,
  title={Diode effect in Josephson junctions with a single magnetic atom},
  author={Trahms, Martina and Melischek, Larissa and Steiner, Jacob F and Mahendru, Bharti and Tamir, Idan and Bogdanoff, Nils and Peters, Olof and Reecht, Ga{\"e}l and Winkelmann, Clemens B and von Oppen, Felix and others},
  journal={Nature},
  volume={615},
  number={7953},
  pages={628--633},
  year={2023},
  publisher={Nature Publishing Group UK London}
}

@article{HuSunLaw2023josephsondiode,
  title={Josephson diode effect induced by valley polarization in twisted bilayer graphene},
  author={Hu, Jin-Xin and Sun, Zi-Ting and Xie, Ying-Ming and Law, KT},
  journal={Physical review letters},
  volume={130},
  number={26},
  pages={266003},
  year={2023},
  publisher={APS}
}

@article{BCS1957,
  title={Theory of superconductivity},
  author={Bardeen, John and Cooper, Leon N and Schrieffer, John Robert},
  journal={Physical review},
  volume={108},
  number={5},
  pages={1175},
  year={1957},
  publisher={APS}
}

@article{Bardeen1962critical,
  title={Critical fields and currents in superconductors},
  author={Bardeen, John},
  journal={Reviews of modern physics},
  volume={34},
  number={4},
  pages={667},
  year={1962},
  publisher={APS}
}

@book{SchriefferSCtextbook,
  title={Theory of superconductivity},
  author={Schrieffer, J Robert},
  year={2018},
  publisher={CRC press}
}

@book{TinkhamSCtextbook,
  title={Introduction to superconductivity},
  author={Tinkham, Michael},
  year={2004},
  publisher={Courier Corporation}
}

@book{deGennesSCtextbook,
  title={Superconductivity of metals and alloys},
  author={De Gennes, Pierre-Gilles},
  year={2018},
  publisher={CRC press}
}

@article{FF1964,
  title={Superconductivity in a strong spin-exchange field},
  author={Fulde, Peter and Ferrell, Richard A},
  journal={Physical Review},
  volume={135},
  number={3A},
  pages={A550},
  year={1964},
  publisher={APS}
}

@article{LO1965,
  title={Zh. {\'e} ksp. teor. fiz. 47, 1136 1964 sov. phys},
  author={Larkin, AI and Ovchinnikov, Yu N},
  journal={JETP},
  volume={20},
  pages={762},
  year={1965}
}

@book{Altland2010textbook,
  title={Condensed matter field theory},
  author={Altland, Alexander and Simons, Ben D},
  year={2010},
  publisher={Cambridge university press}
}

@article{Kleinert1978collectiveQF,
  title={Collective quantum fields},
  author={Kleinert, Hagen},
  journal={Fortschritte der Physik},
  volume={26},
  number={11-12},
  pages={565--671},
  year={1978},
  publisher={Wiley Online Library}
}

@article{Mineev1994helicalSC,
  title={Helical phases in superconductors},
  author={Mineev, VP and Samokhin, KV},
  journal={Zh. Eksp. Teor. Fiz},
  volume={105},
  pages={747--763},
  year={1994}
}

@book{Leggett2006Qliquid,
  title={Quantum liquids: Bose condensation and Cooper pairing in condensed-matter systems},
  author={Leggett, Anthony J},
  year={2006},
  publisher={Oxford university press}
}

@article{LiuDean2022crossover,
  title={Crossover between strongly coupled and weakly coupled exciton superfluids},
  author={Liu, Xiaomeng and Li, JIA and Watanabe, Kenji and Taniguchi, Takashi and Hone, James and Halperin, Bertrand I and Kim, Philip and Dean, Cory R},
  journal={Science},
  volume={375},
  number={6577},
  pages={205--209},
  year={2022},
  publisher={American Association for the Advancement of Science}
}

@article{ShaoDai2024Xbreakdown,
  title={Electrical breakdown of excitonic insulators},
  author={Shao, Yuelin and Dai, Xi},
  journal={Physical Review X},
  volume={14},
  number={2},
  pages={021047},
  year={2024},
  publisher={APS}
}

@article{VarleyLee2016imbalancedeh,
  title={Structure of exciton condensates in imbalanced electron-hole bilayers},
  author={Varley, JR and Lee, DKK},
  journal={Physical Review B},
  volume={94},
  number={17},
  pages={174519},
  year={2016},
  publisher={APS}
}

@article{Seradjeh2012imbalancedeh,
  title={Topological exciton condensate of imbalanced electrons and holes},
  author={Seradjeh, Babak},
  journal={Physical Review B—Condensed Matter and Materials Physics},
  volume={85},
  number={23},
  pages={235146},
  year={2012},
  publisher={APS}
}

@article{DaiFu2024trion,
  title={Strong-coupling phases of trions and excitons in electron-hole bilayers at commensurate densities},
  author={Dai, David D and Fu, Liang},
  journal={Physical Review Letters},
  volume={132},
  number={19},
  pages={196202},
  year={2024},
  publisher={APS}
}

@article{WangDu2023XTO,
  title={Excitonic topological order in imbalanced electron--hole bilayers},
  author={Wang, Rui and Sedrakyan, Tigran A and Wang, Baigeng and Du, Lingjie and Du, Rui-Rui},
  journal={Nature},
  volume={619},
  number={7968},
  pages={57--62},
  year={2023},
  publisher={Nature Publishing Group UK London}
}

@article{KumarSenthil2024XDW,
  title={Unconventional superconductivity mediated by exciton density wave fluctuations},
  author={Kumar, Ajesh and Patri, Adarsh S and Senthil, T},
  journal={arXiv preprint arXiv:2410.09148},
  year={2024}
}

@article{Chen2005bcsbec,
  title={BCS--BEC crossover: From high temperature superconductors to ultracold superfluids},
  author={Chen, Qijin and Stajic, Jelena and Tan, Shina and Levin, Kathryn},
  journal={Physics Reports},
  volume={412},
  number={1},
  pages={1--88},
  year={2005},
  publisher={Elsevier}
}

@book{Volovik2003universeHe,
  title={The universe in a helium droplet},
  author={Volovik, Grigory E},
  volume={117},
  year={2003},
  publisher={OUP Oxford}
}

@article{Lu2017janus,
  title={Janus monolayers of transition metal dichalcogenides},
  author={Lu, Ang-Yu and Zhu, Hanyu and Xiao, Jun and Chuu, Chih-Piao and Han, Yimo and Chiu, Ming-Hui and Cheng, Chia-Chin and Yang, Chih-Wen and Wei, Kung-Hwa and Yang, Yiming and others},
  journal={Nature nanotechnology},
  volume={12},
  number={8},
  pages={744--749},
  year={2017},
  publisher={Nature Publishing Group UK London}
}

@article{Tang2022Janus,
  title={2D Janus transition metal dichalcogenides: Properties and applications},
  author={Tang, Xiao and Kou, Liangzhi},
  journal={physica status solidi (b)},
  volume={259},
  number={4},
  pages={2100562},
  year={2022},
  publisher={Wiley Online Library}
}

@article{MaruyamaSigrist2012sandwich,
  title={Locally non-centrosymmetric superconductivity in multilayer systems},
  author={Maruyama, Daisuke and Sigrist, Manfred and Yanase, Youichi},
  journal={Journal of the Physical Society of Japan},
  volume={81},
  number={3},
  pages={034702},
  year={2012},
  publisher={The Physical Society of Japan}
}

@article{Narozhny2016dragreview,
  title={Coulomb drag},
  author={Narozhny, BN and Levchenko, A},
  journal={Reviews of Modern Physics},
  volume={88},
  number={2},
  pages={025003},
  year={2016},
  publisher={APS}
}

@article{Jauho1993Coulombdrag,
  title={Coulomb drag between parallel two-dimensional electron systems},
  author={Jauho, Antti-Pekka and Smith, Henrik},
  journal={Physical Review B},
  volume={47},
  number={8},
  pages={4420},
  year={1993},
  publisher={APS}
}

@article{Rojo1999Electrondrag,
  title={Electron-drag effects in coupled electron systems},
  author={Rojo, Alberto G},
  journal={Journal of Physics: Condensed Matter},
  volume={11},
  number={5},
  pages={R31},
  year={1999},
  publisher={IOP Publishing}
}

@article{FuHe2025NRCD,
  title={Non-reciprocal Coulomb drag between Chern insulators},
  author={Fu, Yu and Huang, Yu and He, Qing Lin},
  journal={Nature Communications},
  volume={16},
  number={1},
  pages={3058},
  year={2025},
  publisher={Nature Publishing Group UK London}
}

@article{ZverevichLevchenko2025NRCD,
  title={Nonreciprocal Coulomb drag in electron bilayers},
  author={Zverevich, Dmitry and Levchenko, Alex},
  journal={arXiv preprint arXiv:2504.08679},
  year={2025}
}

@article{Rikken2001electricalMCA,
  title={Electrical magnetochiral anisotropy},
  author={Rikken, GLJA and F{\"o}lling, J and Wyder, P},
  journal={Physical review letters},
  volume={87},
  number={23},
  pages={236602},
  year={2001},
  publisher={APS}
}

@article{Ideue2017bulkpolarMCA,
  title={Bulk rectification effect in a polar semiconductor},
  author={Ideue, T and Hamamoto, K and Koshikawa, S and Ezawa, M and Shimizu, S and Kaneko, Y and Tokura, Y and Nagaosa, N and Iwasa, Y},
  journal={Nature Physics},
  volume={13},
  number={6},
  pages={578--583},
  year={2017},
  publisher={Nature Publishing Group UK London}
}

@article{OuyangShi2025trionnews,
  title={Quantum trions in equilibrium},
  author={Ouyang, Tianyi and Shi, Su-Fei},
  journal={Science},
  volume={390},
  number={6770},
  pages={241--241},
  year={2025},
  publisher={American Association for the Advancement of Science}
}

@article{QiWang2025trion,
  title={Electrically controlled interlayer trion fluid in electron-hole bilayers},
  author={Qi, Ruishi and Li, Qize and Zhang, Zuocheng and Chen, Sudi and Xie, Jingxu and Ou, Yunbo and Cui, Zhiyuan and Dai, David D and Joe, Andrew Y and Taniguchi, Takashi and others},
  journal={Science},
  volume={390},
  number={6770},
  pages={299--303},
  year={2025},
  publisher={American Association for the Advancement of Science}
}

@article{NguyenShanMak2025trion,
  title={An equilibrium trion liquid in atomic double layers},
  author={Nguyen, Phuong X and Chaturvedi, Raghav and Ma, Liguo and Knuppel, Patrick and Watanabe, Kenji and Taniguchi, Takashi and Mak, Kin Fai and Shan, Jie},
  journal={Science},
  volume={390},
  number={6770},
  pages={304--307},
  year={2025},
  publisher={American Association for the Advancement of Science}
}

@article{ZerbaKnap2024mixturepwave,
  title={Realizing topological superconductivity in tunable Bose-Fermi mixtures with transition metal dichalcogenide heterostructures},
  author={Zerba, Caterina and Kuhlenkamp, Clemens and Imamo{\u{g}}lu, Ata{\c{c}} and Knap, Michael},
  journal={Physical Review Letters},
  volume={133},
  number={5},
  pages={056902},
  year={2024},
  publisher={APS}
}

@article{Yu2023Janustransport,
  title={Current-induced spin polarization in Janus WSSe monolayer},
  author={Yu, Sheng-Bin and Sun, Shu-Han and Zhou, Ma and Zhang, Dong and Chang, Kai},
  journal={Physical Review B},
  volume={107},
  number={12},
  pages={125426},
  year={2023},
  publisher={APS}
}

@article{XiaoYao2012TMDvalley,
  title={Coupled spin and valley physics in monolayers of MoS 2 and other group-VI dichalcogenides},
  author={Xiao, Di and Liu, Gui-Bin and Feng, Wanxiang and Xu, Xiaodong and Yao, Wang},
  journal={Physical review letters},
  volume={108},
  number={19},
  pages={196802},
  year={2012},
  publisher={APS}
}

@article{Mak2010MoS2,
  title={Atomically thin MoS 2: a new direct-gap semiconductor},
  author={Mak, Kin Fai and Lee, Changgu and Hone, James and Shan, Jie and Heinz, Tony F},
  journal={Physical review letters},
  volume={105},
  number={13},
  pages={136805},
  year={2010},
  publisher={APS}
}

@article{Shankar1994RGfermion,
  title={Renormalization-group approach to interacting fermions},
  author={Shankar, Rev},
  journal={Reviews of Modern Physics},
  volume={66},
  number={1},
  pages={129},
  year={1994},
  publisher={APS}
}

@article{Zollner2019proximity,
  title={Proximity exchange effects in MoSe 2 and WSe 2 heterostructures with CrI 3: Twist angle, layer, and gate dependence},
  author={Zollner, Klaus and Faria Junior, Paulo E and Fabian, Jaroslav},
  journal={Physical Review B},
  volume={100},
  number={8},
  pages={085128},
  year={2019},
  publisher={APS}
}

@article{Ren2023measurement,
  title={Measurement of the conduction band spin-orbit splitting in WSe 2 and WS 2 monolayers},
  author={Ren, Lei and Robert, Cedric and Dery, Hanan and He, Minhao and Li, Pengke and Van Tuan, Dinh and Renucci, Pierre and Lagarde, Delphine and Taniguchi, Takashi and Watanabe, Kenji and others},
  journal={Physical Review B},
  volume={107},
  number={24},
  pages={245407},
  year={2023},
  publisher={APS}
}

@article{Ge2022enhanced,
  title={Enhanced valley splitting of WSe2 in twisted van der Waals WSe2/CrI3 heterostructures},
  author={Ge, Mei and Wang, Han and Wu, Jizheng and Si, Chen and Zhang, Junfeng and Zhang, Shengbai},
  journal={npj Computational Materials},
  volume={8},
  number={1},
  pages={32},
  year={2022},
  publisher={Nature Publishing Group UK London}
}

@misc{supplemental_material,
    title        = {See Supplemental Material at [url] for detailed derivations of the frictional Coulomb drag, discussions on material realization and some other aspects of the work. The supplemental material includes Refs. \cite{Yu2023Janustransport,Zollner2019proximity,Ge2022enhanced,Ren2023measurement}},
    author       = {Jun-Xiao Hui and Qing-Dong Jiang},
    howpublished = {},
    note         = {},
    year         = 2026,
}

@article{Laine2016TFT,
  title={Basics of thermal field theory},
  author={Laine, Mikko and Vuorinen, Aleksi},
  journal={Lect. Notes Phys},
  volume={925},
  number={1},
  pages={1701--01554},
  year={2016},
  publisher={Springer}
}

@article{shao2025electromagnetic,
  title={Electromagnetic responses of bilayer excitonic insulators: from exciton London equations to dipole and inverse dipole Hall effects},
  author={Shao, Yuelin and Shi, Hao and Dai, Xi},
  journal={arXiv preprint arXiv:2509.02142},
  year={2025}
}

@data{DVN/NYXL70_2026,
  author = {Hui, Jun-Xiao and Jiang, Qing-Dong},
  publisher = {Harvard Dataverse},
  title = {Data for ``Nonreciprocal Perfect Coulomb Drag in Electron–Hole Bilayers: Coherent Exciton Superflow as a Diode''},
  year = {2026},
  version = {V1},
  doi = {10.7910/DVN/NYXL70},
  url = {https://doi.org/10.7910/DVN/NYXL70},
  note = {Available at: \url{https://doi.org/10.7910/DVN/NYXL70}}
}

\end{document}



\title{Supplemental Materials for ``Nonreciprocal Perfect Coulomb Drag in Electron–Hole Bilayers: Coherent Exciton Superflow as a Diode"}


\author{Jun-Xiao Hui}
\affiliation{Tsung-Dao Lee Institute, Shanghai Jiao Tong University, Shanghai 200240, China}
\affiliation{School of Physics and Astronomy, Shanghai Jiao Tong University, Shanghai 200240, China}

\author{Qing-Dong Jiang}
\email{qingdong.jiang@sjtu.edu.cn}
\affiliation{Tsung-Dao Lee Institute, Shanghai Jiao Tong University, Shanghai 200240, China}
\affiliation{School of Physics and Astronomy, Shanghai Jiao Tong University, Shanghai 200240, China}
\affiliation{Shanghai Branch, Hefei National Laboratory, Shanghai 201315, China}

\date{\today}


\maketitle
\appendix



\section{Supplemental Materials}

\subsection{S-\Rmnum{1}. Rashba--Zeeman--Ising model and choice of parameters}\label{sec:RZI model}

{\bf  Rashba--Zeeman--Ising model.} In this section, we connect the effective model used in the main text to realistic van der Waals heterostructures and provide parameter estimates relevant for experiments. The goal is to clarify how the band structure, spin--orbit coupling, Zeeman fields, gate voltages, and device geometry enter the model, and to show that the predicted nonreciprocal Coulomb drag effect can be reached within experimentally accessible regimes.

The theoretical calculations in the main text are based on the following effective Hamiltonians for electrons in the top layer and holes in the bottom layer:
\be
H_{t,e} &=& (\frac{k^2}{2m_e}-\mu_e) + \lambda_{R}(k_x\sigma_y-k_y\sigma_x) + B\sigma_y + J_t\sigma_z \label{eq:teHamiltonian} \\
H_{b,h} &=& (\frac{k^2}{2m_h}-\mu_h) + B\sigma_y + J_b\sigma_z \label{eq:bhHamiltonian}
\ee
Here $\lambda_R$ denotes the Rashba spin--orbit coupling, which originates from out-of-plane inversion-symmetry breaking. The term $B\sigma_y$ describes an in-plane Zeeman field. The parameters $J_t$ and $J_b$ denote out-of-plane spin splittings in the top and bottom layers, respectively. In transition-metal dichalcogenide monolayers, these terms naturally arise from Ising spin--orbit coupling associated with in-plane inversion-symmetry breaking. 
Directly diagonalization of the Hamiltonian yields the electron and hole dispersions: 
\be
\xi_{e,\pm} = \frac{k^2}{2m_e} - \mu_e \pm \lambda_R\sqrt{(k_x+B/\lambda_R)^2+k_y^2+(J_t/\lambda_R)^2},\, \xi_{h,\pm} = \frac{k^2}{2m_h}-\mu_h \pm \sqrt{B^2+J_b^2} \label{eq:ehdispersionsfull}
\ee

The top and bottom chemical potentials can be tuned independently by gate voltages. Let $\Delta_t$ denote the energy difference between the Fermi level and the conduction-band minimum of the top layer before gating, and let $\Delta_b$ denote the energy difference between the Fermi level and the valence-band maximum of the bottom layer before gating. Then the gate-tuned chemical potentials may be written as $\mu_e=eV_t-\Delta_t,~\mu_h=-eV_b-\Delta_b$. By tuning $V_t$ and $V_b$, the electron and hole densities can be controlled independently.

In the regime where the Fermi level intersects only one spin-polarized conduction band in the top layer and one spin-polarized valence band in the bottom layer, the low-energy physics is governed by the lower branches $\xi_{e,-}$ and $\xi_{h,-}$. This band isolation justifies the effective single-component model used in the main text and allows us to perform the mean-field calculation for finite-momentum exciton condensation.
\\\\

{\bf Parameter values used in the main text.} For the numerical results shown in the main text, we used the representative parameter set
\begin{equation}
m_e=m_h=m_*=0.4m_0,
\qquad
\lambda_R=0.80~{\rm eV\AA},
\qquad
J_t=J_b=17.2~{\rm meV},
\label{eq}
\end{equation}
where $m_0\simeq 9.11\times 10^{-31}~{\rm kg}$ is the free-electron mass.

We use the convenient numerical convention
$m_*=\hbar=k_B T_c=1$
with $T_c=10~{\rm K}$. This gives the energy and length units
\begin{equation}
E_u=k_B T_c\simeq 0.86~{\rm meV},
\qquad
L_u=
\frac{\hbar}{\sqrt{k_B T_c m_*}}
\simeq
93.4~{\rm nm}.
\end{equation}
In the main-text calculations, we take
\begin{equation}
\mu_e\simeq 18.0~{\rm meV},
\qquad
\mu_h\simeq 17.5~{\rm meV}.
\end{equation}
For the in-plane Zeeman field, we use $B=3E_u\simeq 2.58~{\rm meV}$ in Figs.~3a--3c and Fig.~4a of the main text. We then vary $B$ from $-20E_u\simeq -17.2~{\rm meV}$ to $20E_u\simeq 17.2~{\rm meV}$ in Figs.~3d and 4b.

The Rashba coupling $\lambda_R=0.80~{\rm eV\AA}$ is of the same order as values reported for Janus TMD monolayers such as WSSe~\cite{Yu2023Janustransport}. An in-plane Zeeman splitting of $1~{\rm meV}$ corresponds roughly to an external magnetic field of order $8~{\rm T}$. Thus $B=3E_u\simeq 2.58~{\rm meV}$ is within the range of high-field experiments, while the largest value $B=20E_u\simeq 17.2~{\rm meV}$ should be viewed as a theoretical upper range rather than an immediately accessible experimental value.

The out-of-plane spin splittings $J_t$ and $J_b$ do not have to be generated by an external magnetic field. In TMD monolayers such as MoSe$_2$ and WSe$_2$, large out-of-plane spin splittings arise naturally from intrinsic Ising spin--orbit coupling~\cite{XiaoYao2012TMDvalley}. A complementary route is magnetic proximity engineering, in which adjacent magnetic layers induce sizable spin-exchange fields, typically on the meV scale~\cite{Zollner2019proximity,Ge2022enhanced,Ren2023measurement}.

These considerations show that the representative parameters used in the main text are physically motivated and lie in the range of values relevant to TMD-based van der Waals materials. Nevertheless, quantitative comparison with a specific experimental device requires extracting material parameters case by case. We therefore next discuss a concrete heterostructure in which the model parameters can be directly connected to realistic materials.

\subsection{S-\Rmnum{2}. Proposed WSSe/hBN/WSe$_2$ heterostructure}
We propose a WSSe/hBN/WSe$_2$ van der Waals heterostructure  (Fig. 1)as a concrete platform for realizing the model in Eqns. \eqref{eq:teHamiltonian} \& \eqref{eq:bhHamiltonian}. The device consists of a monolayer WSSe electron layer and a monolayer WSe$_2$ hole layer separated by a thin insulating hBN spacer. Top and bottom gates independently tune the carrier densities through gate voltages $V_t$ and $V_b$. The conduction band of WSSe supplies electron-like carriers, while the valence band of WSe$_2$ supplies hole-like carriers.
\begin{figure}[htp]
\includegraphics[width=0.4\linewidth]{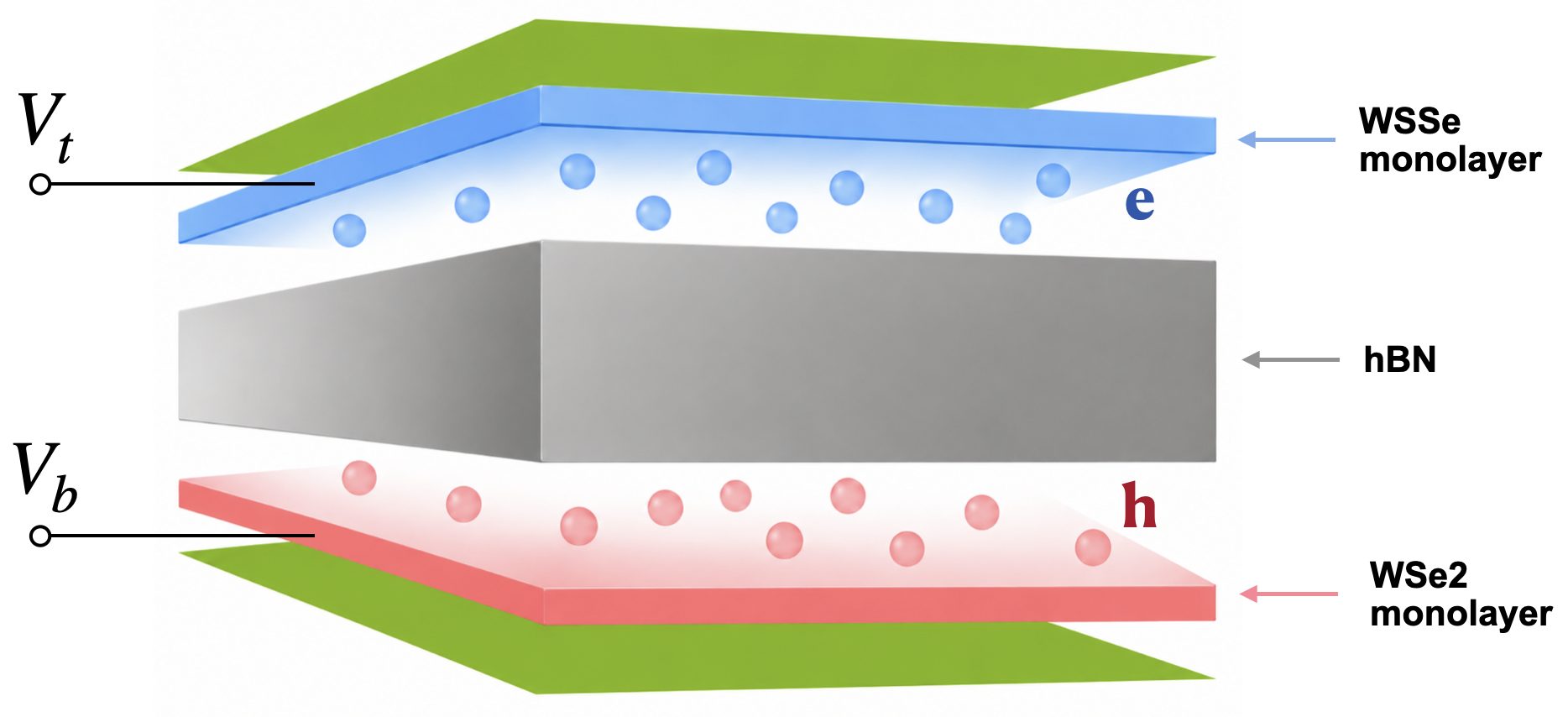}
    \caption{The VdW heterostrcture WSSe/hBN/WSe$_2$ is encapsulated by independently tunable top and bottom gates with gate voltage $V_t$ and $V_b$. The top (blue) layer is a monolayer WSSe and the bottom (red) layer is a monolaye WSe$_2$. The two TMD layers are separated by a thin insulating hBN spacer (gray), typically a few nanometers thick, which enables strong interlayer Coulomb interactions while suppressing direct tunneling. }
    \label{fig:replysetup}
\end{figure}
The key advantage of WSSe is that it breaks inversion symmetry both in plane and out of plane. Consequently, it naturally supports both Ising and Rashba spin--orbit couplings. For electrons in the WSSe conduction band, the Rashba coupling is approximately $\lambda_R\simeq 0.3187~{\rm eV\AA}$, and the Ising spin splitting is approximately $J_t\simeq 31.8~{\rm meV}$~\cite{Yu2023Janustransport}.
For the WSe$_2$ valence band, the intrinsic Ising splitting is much larger, with a total spin splitting of order $0.46~{\rm eV}$~\cite{XiaoYao2012TMDvalley}. Thus one spin-polarized valence band can be isolated over a broad range of carrier densities.

Using reported material parameters, we estimate
\begin{equation}
m_e \approx 0.4m_0,
\qquad
m_h \approx 0.5m_0,
\qquad
\lambda_R \approx 0.32~{\rm eV\AA},
\qquad
J_t = 31.8~{\rm meV},
\qquad
J_b = 230~{\rm meV}.
\label{eq}
\end{equation}
These values provide a realistic parameter set for implementing the effective Hamiltonians in in Eqns. \eqref{eq:teHamiltonian} \& \eqref{eq:bhHamiltonian}. Notably, both WSSe and WSe$_2$ possess two valleys related by time-reversal symmetry. In general, this can lead to a two-component exciton condensate. However, as discussed in Ref.~\cite{WuMacdonald2015theory}, a single-component condensate can be energetically selected in suitable regions of the phase diagram. Our calculation focuses on such a single-component regime, where the low-energy physics is captured by the model used in the main text.

\subsection{S-\Rmnum{3}. Numerical estimate for the realistic device}

We now recalculate free energy landscape in momentum space and coherent exciton diode efficiency using the realistic parameter set in Eq.~\eqref{eq}. We choose
\begin{equation}
B=5.0E_u\simeq 4.3~{\rm meV},
\end{equation}
corresponding to an external magnetic field of approximately $36~{\rm T}$. Although this is a large magnetic field, it is within the range of modern high-field transport experiments. We also choose
\begin{equation}
\mu_e+J_t=43E_u\simeq 37.0~{\rm meV},
\qquad
\mu_h+J_b=36.2E_u\simeq 31.1~{\rm meV},
\end{equation}
so that the electron and hole Fermi surfaces have comparable sizes.

\begin{figure}[h]
\centering
\includegraphics[width=0.6\linewidth]{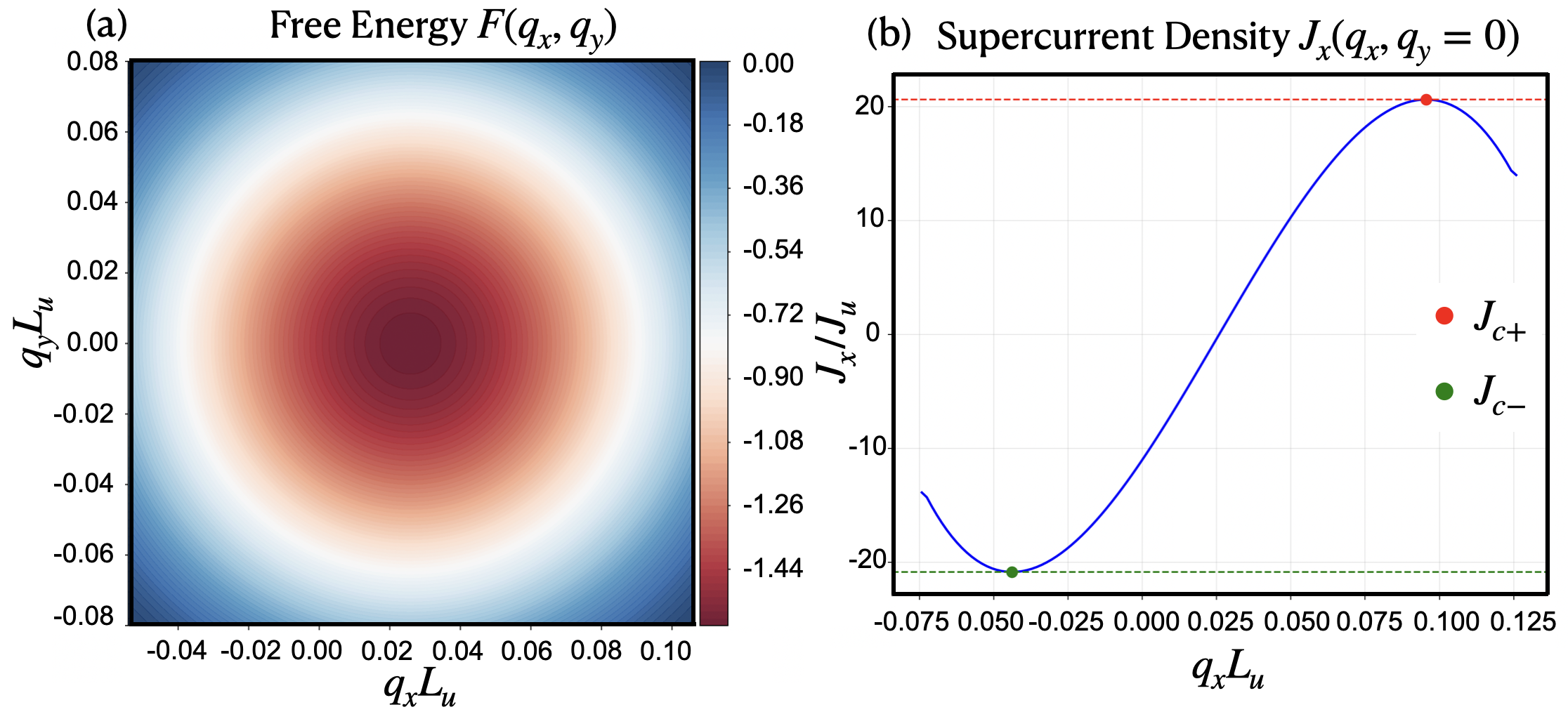}
\caption{
Numerical results for the realistic WSSe/hBN/WSe$_2$ heterostructure. a, Free-energy landscape of the finite-momentum exciton condensate. b, Supercurrent $J_x$ as a function of condensate momentum $q_x$. The critical currents are $J_{x+}\approx 20.640J_u$ and $J_{x-}\approx -20.838J_u$, giving a diode efficiency $\eta=\left(|J_{x+}|-|J_{x-}|\right)/\left(|J_{x+}|+|J_{x-}|\right)\approx -0.476\% $.
}
\label{fig}
\end{figure}

The resulting free-energy landscape and supercurrent are shown in Fig.~\ref{fig}. We find a diode efficiency of approximately
\begin{equation}
|\eta|\simeq 0.5
\end{equation}
For a device with effective width $1~\mu{\rm m}$ along the $y$ direction, the corresponding critical currents are
\begin{equation}
I_{c+}\approx 46.234~\mu{\rm A},
\qquad
I_{c-}\approx -46.677~\mu{\rm A}.
\end{equation}
The nonreciprocal critical-current difference is therefore
\begin{equation}
\delta I_c=
\frac{I_{c+}+I_{c-}}{2}
\approx
-221.5~{\rm nA}.
\end{equation}
This signal is well within the resolution of present-day transport measurements.

Compared with the idealized parameter set used in the main text, the diode efficiency in this realistic device is smaller. The main reason is that the Ising splitting $J_t$ of the top-layer electrons is larger in WSSe. Since the electron dispersion contains the combination
$\pm
\sqrt{
\left(
\lambda_R k_x-B
\right)^2
+
\left(
\lambda_R k_y
\right)^2
+
J_t^2
},
$
a larger $J_t$ suppresses the momentum-space asymmetry induced by the in-plane Zeeman field $B$. This explains why the realistic WSSe/hBN/WSe$_2$ platform gives a smaller diode efficiency than the optimized model studied in the main text.

The above analysis suggests several possible routes to improve the diode effect experimentally. One may search for Janus TMD monolayers or related two-dimensional materials with strong Rashba coupling but weaker Ising splitting in the electron layer. Alternatively, one may engineer a tunable spin splitting by proximity coupling the electron layer to a magnetic material~\cite{Ge2022enhanced,Ren2023measurement}. Such proximity engineering could allow one to optimize $J_t$ and enhance the nonreciprocal Coulomb drag response.

Overall, this realistic-material analysis shows that the exciton diode effect is not limited to an idealized model. A measurable nonreciprocal critical-current difference can arise in experimentally accessible TMD-based van der Waals heterostructures, and further material optimization may lead to even larger diode efficiencies.

\subsection{S-\Rmnum{4}. Functional Approach to Bilayer Exciton Condensate}\label{sec:Functional mean field}

In the model presented in previous Section, we isolate the lower conduction band in the top layer and upper valence band in the bottom layer. Introducing hole creation/annihilation operators $\chi_{k}=c_{-k}^{\dagger}$ in bottom layer, we end up with a two-species fermion system. The effective second-quantized Hamiltonian is:
\be
H = \sum_k(\xi_{e,k}\psi_{k}^{\dagger}\psi_k + \xi_{h,k}\chi_{k}^\dagger \chi_k) + V_{eh} \label{Hdoublelayer}
\ee
Where $\xi_{e,k}=\frac{k^2}{2m_e}-\lambda\sqrt{(k_x+(B/\lambda))^2+k_y^2+(J/\lambda)^2}-E_{Fe},\xi_{h,k}=\frac{k^2}{2m_h}-E_{Fh}$. $E_{Fe},E_{Fh}$ measure the energy difference from Fermi level to band bottoms. The form of $V_{eh}$ is presented in the main text, which can be written as $- (g/\mathcal{V})\sum_{k,k^{\prime}}\psi_{k+q/2}^{\dagger}\chi_{-k+q/2}^{\dagger}\chi_{-k^{\prime}+q/2}\psi_{k^{\prime}+q/2}$ under \s-wave, $q$-momentum pairing approximation. In the functional integral formalism, the action reads:
\be
S[\dagr{\psi},\psi,\dagr{\chi},\chi] = \int d\tau\,\sum_k(\dagr{\psi}_{k}(\partial_\tau+\xi_{e,k})\psi_k + \dagr{\chi}_{k}((\partial_\tau+\xi_{h,k}))\chi_k) - \frac{g}{\mathcal{V}}\sum_{k,k^{\prime}}\psi_{k+q/2}^{\dagger}\chi_{-k+q/2}^{\dagger}\chi_{-k^{\prime}+q/2}\psi_{k^{\prime}+q/2}  \label{Sdoublelayer}
\ee
We perform the Hubbard-Stratonovich transformation and extract the effective theory of excitonic order parameter field~\cite{Altland2010textbook,Kleinert1978collectiveQF}: 
\be
\mathcal{Z} &=& \int D(\dagr{\psi},\psi,\dagr{\chi},\chi)\exp\{-S[\dagr{\psi},\psi,\dagr{\chi},\chi]\} \\
            &=& \int D(\bar{\Delta},\Delta)\int D(\dagr{\psi},\psi,\dagr{\chi},\chi)\,\exp\{-\int d\tau\,[\frac{\mathcal{V}}{g}\bar{\Delta}\Delta+\sum_{k}(\dagr{\psi}_{k}(\partial_\tau+\xi_{e,k})\psi_k + \dagr{\chi}_{k}((\partial_\tau+\xi_{h,k}))\chi_k \\
            &-& \bar{\Delta}\chi_{-k+q/2}\psi_{k+q/2}-\dagr{\psi}_{k+q/2}\dagr{\chi}_{-k+q/2}\Delta)]\} \\
            &=& \int D(\bar{\Delta},\Delta)\,\exp\{-\Gamma[\bar{\Delta},\Delta]\}
\ee
Where
\be
\exp\{-\Gamma[\bar{\Delta},\Delta]\} &=& \exp\{-\int d\tau\,\frac{\mathcal{V}}{g}\bar{\Delta}\Delta\}\int D(\dagr{\psi},\psi,\dagr{\chi},\chi)\,\exp\{-\int d\tau\,\sum_{k}\,\begin{pmatrix}
    \dagr{\psi}_{k+q/2} & \chi_{-k+q/2}
\end{pmatrix} \\
                                     &\times&  \begin{pmatrix}
                                         \partial_\tau+\xi_{e,k+q/2} & -\Delta \\
                                         -\bar{\Delta} & \partial_\tau-\xi_{h,-k+q/2}
                                        \end{pmatrix} \begin{pmatrix}
                                            \psi_{k+q/2} \\
                                            \dagr{\chi}_{-k+q/2}
                                        \end{pmatrix}\} \\
                                     &=& \exp\{-\frac{\mathcal{V}\beta}{g}\bar{\Delta}\Delta + \sum_{n,k}\ln[(-i\omega_n+\xi_{e,k+q/2})(-i\omega_n-\xi_{h,-k+q/2})-\bar{\Delta}\Delta]\} \\
                                     &=& \exp\{-\beta F[\bar{\Delta},\Delta]\}
\ee
Where we have neglected temporal fluctuation by assuming $\Delta(\tau)\equiv\Delta$, which can be justified near transition temperature~\cite{Laine2016TFT}. The free energy functional is defined as
\be
F[\bar{\Delta},\Delta] &=& T\Gamma[\bar{\Delta},\Delta] \\
                       &=& \frac{\mathcal{V}}{g}\bar{\Delta}\Delta - T\sum_{n,k}\ln[(-i\omega_n+\xi_{e,k+q/2})(-i\omega_n-\xi_{h,-k+q/2})-\bar{\Delta}\Delta] \label{eq:freeenergyfunctional} \\
                       &=& F_0 + \frac{\mathcal{V}}{g}\bar{\Delta}\Delta + T\sum_{l=1}^{\infty}\frac{1}{l}\sum_{n,k}\frac{1}{(-i\omega_n+\xi_{e,k+q/2})^l(-i\omega_n-\xi_{h,-k+q/2})^l}(\bar{\Delta}\Delta)^l \label{eq:freeenergyfunctionalexpansion}
\ee
Where $F_0=-T\sum_{n,k}\ln[(-i\omega_n+\xi_{e,k+q/2})(-i\omega_n-\xi_{h,-k+q/2})]$ is the free energy of original fermion gases, which one can verify after some simple algebra~\cite{Altland2010textbook}. Near but slightly below critical temperature, $|\Delta|/T \ll 1$, then we can keep the functional to leading relevant order of $(\bar{\Delta},\Delta)$ to obtain Ginzburg-Landau free energy functional:
\be
F_{GL}[\bar{\Delta},\Delta] = \alpha(T,q)(\bar{\Delta}\Delta)+\frac{1}{2}\beta(T,q)(\bar{\Delta}\Delta)^2 \label{eq:GLfreeenergy}
\ee
Where 
\be
\alpha &=& \frac{\mathcal{V}}{g} + T\sum_{n,k}\frac{1}{(-i\omega_n+\xi_{e,k+q/2})(-i\omega_n-\xi_{h,-k+q/2})} \\
\beta  &=& T\sum_{n,k}\frac{1}{(-i\omega_n+\xi_{e,k+q/2})^2(-i\omega_n-\xi_{h,-k+q/2})^2}
\ee
as presented in the main text of the paper. At specific $q$, $\alpha(T<T_c(q))<0$ while $\alpha(T>T_c)>0$, which characterizes the superconducting-like phase transition of the exciton condensate.


\subsection{S-\Rmnum{5}. Kinetic Theory for Frictional Coulomb Drag}
When the double-layer system has not transitioned into an exciton condensate, Coulomb drag still exists, which originates from momentum transfer between the layers due to Coulomb interaction and is called frictional drag~\cite{Narozhny2016dragreview}. Suppose the drag layer is an open circuit, current in drive layer $j_1$ will induce a voltage $V_2$ therein(I-V drag), and drag resistivity $\rho_D=E_2/j_1$ is defined to characterize the strength of the drag. Alternatively, one can short circuit the drag layer, and a current $j_2$ is induced by $j_1$(I-I drag), and drag ratio $\zeta=j_2/j_1$ is defined in this case. The second scenario corresponds to our case. In steady state, both $\rho_D$ and $\zeta$ can be calculated using Boltzmann transport equation~\cite{Narozhny2016dragreview,Jauho1993Coulombdrag,Rojo1999Electrondrag}. Here we present the derivation of $\rho_D$ and $\zeta$ and express them with microscopic parameters. Furthermore, we will try to go to the leading non-linear order and argue the onset of nonreciprocal drag in presence of $\mP,\T$ symmetry breaking.

\subsubsection{Frictional Coulomb drag for fermions with parabolic dispersion}
Consider a 2-dimensional double-layer fermion system with $d$ the layer separation. The dispersion relations are $\xi^{t,b}_k=\frac{k^2}{2m_{t,b}}-\mu_{t,b}$ for charge carrier fermions in the top(drive) and bottom(drag) layers, respectively. In our case of electron-hole bilayer setup, $t=e\,\&\,b=h$, where $e$ represents electron and $h$ represents hole. Yet we keep the following derivations general and apply the final formulae to our setup eventually.
Denote $f_0(\xi)=(e^{\beta\xi}+1)^{-1}$ as the Fermi-Dirac distribution, and $\tilde{f}_0(\xi)=1-f_0(\xi)=f_0(-\xi)$. We denote $e_1,e_2$ as charge of carriers in top and bottom layers, respectively. Focusing on the drag layer, in a homogeneous steady state, the Boltzmann equation reads:
\be
e_2 \tb{E}_2\cdot\partial_{k_2}f(k_2) = (\frac{\partial f(k_2)}{\partial t})_{drag} + (\frac{\partial f(k_2)}{\partial t})_{relaxation} \label{eq:Boltzmanneqn}
\ee
For simplicity, we use $k_1,k_2$ to label momenta of particles in drive and drag layer respectively, and denote $f(k_1)=f(\xi^{t}_{k_1}),f(k_2)=f(\xi^{b}_{k_2})$. For I-V drag, $(\frac{\partial f(k_2)}{\partial t})_{relaxation} = 0$ since bulk of drag layer is in equilibrium; for I-I drag, $\tb{E}_2=0$ due to short circuit condition. Since the I-V drag is studied extensively in literature~\cite{Jauho1993Coulombdrag,Rojo1999Electrondrag}, we focus on the I-I drag. In the steady state of I-I drag, the two layers are both carrying current though spatially homogeneous, so we use drifting solution
\be
f(k_1)=f_0(k_1)-m_t\tb{v}_1\cdot\partial_{k_1}f_0(k_1)\approx f_0(k_1-m_t\tb{v}_1) \label{eq:fk1} \\
f(k_2)=f_0(k_2)-m_b\tb{v}_2\cdot\partial_{k_2}f_0(k_2)\approx f_0(k_2-m_b\tb{v}_2) \label{eq:fk2}
\ee
Where $v_i$ are drift velocity for either layer and $j_i=n_ie_iv_i$, $i=1,2$. Then the relaxation term in drag layer can be captured by quasi-particle lifetime $\tau_2$:
\be
(\frac{\partial f(k_2)}{\partial t})_{relaxation} = -\frac{f(k_2)-f_0(k_2)}{\tau_2} = \frac{m_b\tb{v}_2}{\tau_2}\cdot\partial_{k_2}f_0(k_2) \label{eq:relaxintegral}
\ee
The drag term originates from interlayer scattering events and can be expressed as:
\be
(\frac{\partial f(k_2)}{\partial t})_{drag} &=& -\intk{q}\intk{k_1}\,\mathcal{A}(q)(f(k_1)\tilde{f}(k_1-q)f(k_2)\tilde{f}(k_2+q)-f(k_1-q)\tilde{f}(k_1)f(k_2+q)\tilde{f}(k_2)) \nonumber \\
                                            &\times& \delta(\xi^t_{k_1}+\xi^b_{k_2}-\xi^t_{k_1-q}-\xi^b_{k_2+q}) \label{eq:dragintegral}
\ee
Where $\mathcal{A}(q)=2\pi|U(q)|^2$ according to Fermi's golden rule, and $U(q)$ is matrix element of the screened interlayer Coulomb interaction as discussed in the main text. Now multiply both sides of \eqref{eq:Boltzmanneqn} with $\tb{k}_2$ and integrate over $\tb{k}_2$, namely apply $\intk{k_2}\tb{k}_2$ to both sides, obtaining:
\be
0 &=& -\intk{k_2}\tb{k}_2\frac{f_0(k_2-m_b\tb{v}_2)-f_0(k_2)}{\tau_2} + \intk{k_2}\tb{k}_2(\frac{\partial f(k_2)}{\partial t})_{drag} \\
  &=& -n_2\frac{m_b}{\tau_2}\tb{v}_2 + \intk{k_2}\tb{k}_2(\frac{\partial f(k_2)}{\partial t})_{drag}
\ee
$n_1,n_2$ are carrier densities in top and bottom layers, respectively. The above equation is literally a force balance condition on average --- in drag layer, drag force $F_{drag}$ exerted by drive layer is balanced with frictional force $\tb{f}=-\frac{m_b}{\tau_2}\tb{v}_2$, namely:
\be
n_2(-\frac{m_b}{\tau_2}\tb{v}_2 + \tb{F}_{drag}(\tb{v}_1,\tb{v}_2)) = 0 \label{eq:forcebalance}
\ee
Where
\be
n_2\tb{F}_{drag}(\tb{v}_1,\tb{v}_2) &=& \intk{k_2}\tb{k}_2(\frac{\partial f(k_2)}{\partial t})_{drag} \\
                                    &=& -\intk{k_2}\tb{k}_2\intk{q}\intk{k_1}\,\mathcal{A}(q)(f(k_1)\tilde{f}(k_1-q)f(k_2)\tilde{f}(k_2+q) \nonumber \\
                                            &-& f(k_1-q)\tilde{f}(k_1)f(k_2+q)\tilde{f}(k_2))\delta(\xi^t_{k_1}+\xi^b_{k_2}-\xi^t_{k_1-q}-\xi^b_{k_2+q}) \\ 
                                    &=& -\intk{k_2}\frac{1}{2}(\tb{k}_2-(\tb{k}_2+\tb{q}))\intk{q}\intk{k_1}\,\mathcal{A}(q)(f(k_1)\tilde{f}(k_1-q)f(k_2)\tilde{f}(k_2+q) \nonumber \\
                                            &-& f(k_1-q)\tilde{f}(k_1)f(k_2+q)\tilde{f}(k_2))\delta(\xi^t_{k_1}+\xi^b_{k_2}-\xi^t_{k_1-q}-\xi^b_{k_2+q}) \\
                                    &=& \frac{1}{2}\intk{q}\tb{q}\mathcal{A}(q)\intk{k_1}\intk{k_2}(f(k_1)\tilde{f}(k_1-q)f(k_2)\tilde{f}(k_2+q) \nonumber \\
                                            &-& f(k_1-q)\tilde{f}(k_1)f(k_2+q)\tilde{f}(k_2))\delta(\xi^t_{k_1}+\xi^b_{k_2}-\xi^t_{k_1-q}-\xi^b_{k_2+q}) \label{eq:dragforce1}
\ee
The substitution $k_2\rightarrow(k_2-(k_2+q))/2$ only holds when $f(k_2)=f(-k_2)$ symmetry is preserved. Denote group velocity $\tb{v}(k_{1,2})=\partial_{k_{1,2}}\xi^{t,b}_{k_{1,2}}$, we can further express \eqref{eq:fk1},\eqref{eq:fk2} as:
\be
f(k_1) = f_0(k_1)(1+\beta m_t\tb{v}_1\cdot \tb{v}(k_1)\tilde{f}_0(k_1)) = f_0(k_1)(1+\beta \tb{v}_1\cdot \tb{k}_1\tilde{f}_0(k_1)) \label{eq:fk1'} \\
f(k_2) = f_0(k_2)(1+\beta m_b\tb{v}_2\cdot \tb{v}(k_2)f_0(k_2)) = f_0(k_2)(1+\beta \tb{v}_2\cdot \tb{k}_2f_0(k_2)) \label{eq:fk2'}
\ee
Along with detailed balance condition:
\be
f_0(\epsilon_1)\tilde{f}_0(\epsilon_2)f_0(\epsilon_3)\tilde{f}_0(\epsilon_4) = f_0(\epsilon_2)\tilde{f}_0(\epsilon_1)f_0(\epsilon_4)\tilde{f}_0(\epsilon_3)\quad if \quad \epsilon_1+\epsilon_3=\epsilon_2+\epsilon_4
\ee
\eqref{eq:dragforce1} can be further expressed as:
\be
n_2\tb{F}_{drag}(\tb{v}_1,\tb{v}_2) &=& \frac{1}{2}\intk{q}\tb{q}\mathcal{A}(q)\intk{k_1}\intk{k_2}f_0(k_1)\tilde{f}_0(k_1-q)f_0(k_2)\tilde{f}_0(k_2+q)[(1+\beta \tb{v}_1\cdot \tb{k}_1\tilde{f}_0(k_1)) \nonumber \\
                                    &\times& (1-\beta \tb{v}_1\cdot (\tb{k}_1-\tb{q})f_0(k_1-q))(1+\beta \tb{v}_2\cdot \tb{k}_2\tilde{f}_0(k_2))(1-\beta \tb{v}_2\cdot (\tb{k}_2+\tb{q})f_0(k_2+q)) \nonumber \\
                                    &-& (1+\beta \tb{v}_1\cdot (\tb{k}_1-\tb{q})\tilde{f}_0(k_1-q))(1-\beta \tb{v}_1\cdot\tb{k}_1f_0(k_1))(1+\beta \tb{v}_2\cdot (\tb{k}_2+\tb{q})\tilde{f}_0(k_2+q)) \nonumber \\
                                    &\times& (1-\beta \tb{v}_2\cdot\tb{k}_2 f_0(k_2))]\delta(\xi^t_{k_1}+\xi^b_{k_2}-\xi^t_{k_1-q}-\xi^b_{k_2+q}) \label{eq:dragforce2}
\ee
From \eqref{eq:dragforce2}, we can expand the drag force to both linear and quadratic order with respect to $v_1,v_2$, namely $j_1,j_2$. Without loss of generality, suppose the drag is in \x direction, at linear order we obtain:
\be
F_{drag}(v_1,v_2) &=& \frac{m_t}{\tau_{D1}}v_1 -\frac{m_b}{\tau_{D2}}v_2 = \gamma_D(v_1-v_2) \label{eq:dragforcefinal}
\ee
Where
\be
\gamma_D = \frac{\beta}{16\pi^2n_2}\intk{q}|q|^2\mathcal{A}(q)\int d\omega\,\frac{\Im\chi_1(q,\omega)\Im\chi_2(q,\omega)}{\sinh^2(\beta\omega/2)} \label{eq:gammaD}
\ee
\be
\frac{1}{\tau_{D1}} = \frac{\gamma_D}{m_t},\quad \frac{1}{\tau_{D2}} = \frac{\gamma_D}{m_b} \label{eq:tauD}
\ee
When the dispersions are parabolic, velocities have the exact meaning as parameters of the Galilean boost. Hence $F_{drag}(v_1,v_2)=F_{drag}(v_1-v_2)$ and vanishes when $v_1=v_2$, which ensures the vanishing of drag force in absence of relative motion. $\chi_{1,2}(q,\omega)=-\int\frac{d^2k}{(2\pi)^2}\frac{f_0(\xi^{t,b}_{k})-f_0(\xi^{t,b}_{k+q})}{\omega-\xi^{t,b}_{k}+\xi^{t,b}_{k+q}+i0^+}$ are susceptibilities of the fermion gases. \\
\newpara Combine \eqref{eq:forcebalance} and \eqref{eq:dragforcefinal}, at linear order, we identify the drag ratio
\be
\zeta = \frac{v_2}{v_1}=\frac{1}{1+\tau_{D2}/\tau_2} \label{eq:dragratio}
\ee
\newpara One can expand \eqref{eq:dragforce2} to quadratic order, and find that all quadratic terms vanish! Symmetry arguments imply the vanishing before calculation: quadratic terms lead to nonreciprocal drag effect, which has no chance to exist since our fermion dispersions are both isotropic. When anisotropy is introduced in fermion dispersions, one would expect non-vanishing quadratic-order terms, which will be discussed subsequently.

\subsubsection{Frictional Coulomb drag with general drive layer dispersion}
Now suppose the drive layer does not have perfect parabolic fermion dispersion any more, e.g. the lower branch in \eqref{eq:ehdispersionsfull}: $\xi^t_k=\frac{k^2}{2m_e} - \lambda_R\sqrt{(k_x+B/\lambda_R)^2+k_y^2+(J/\lambda_R)^2} - \mu_e$, which breaks both inversion and time reversal symmetry. Suppose the drag layer still has parabolic dispersion in order to use the trick in \eqref{eq:dragforce1}, and without loss of generality we perform the drag experiment in x direction, a similar derivation leads to the result:
\be
n_2 F_{drag}(v_1,v_2) &=& \frac{1}{2}\intk{k_1}\intk{k_2}\intk{q}q_x\mathcal{A}(q)f_0(k_1)\tilde{f}_0(k_1-q)f_0(k_2)\tilde{f}_0(k_2+q)[(1+\beta\bar{m}^x_t v_1 v_x(k_1)\tilde{f}_0(k_1)) \nonumber \\
                                    &\times& (1-\beta\bar{m}^x_t v_1 v_x(k_1-q)f_0(k_1-q))(1+\beta v_2 k_{2x}\tilde{f}_0(k_2))(1-\beta v_2(k_{2x}+q_x)f_0(k_2+q)) \nonumber \\
                                    &-& (1+\beta \bar{m}^x_t v_1 v_x(k_1-q)\tilde{f}_0(k_1-q))(1-\beta \bar{m}^x_t v_1 v_x(k_1)f_0(k_1))(1+\beta v_2 (k_{2x}+q_x)\tilde{f}_0(k_2+q)) \nonumber \\
                                    &\times& (1-\beta v_2 k_{2x} f_0(k_2))]\delta(\xi^t_{k_1}+\xi^b_{k_2}-\xi^t_{k_1-q}-\xi^b_{k_2+q}) \label{eq:generaldragforcefull}
\ee
Where $(1/\bar{m}^x_t)=(n_1)^{-1}\intk{k_1}\partial^2_{k_{1x}}\xi^t_{k_1}f_0(k_1)$ is the average effective mass.To the quadratic order of $v_1,v_2$, we obtain
\be
F_{drag}(v_1,v_2) = \frac{\bar{m}^x_t}{\tau_{D1}}v_1-\frac{m_b}{\tau_{D2}}v_2+\kappa_{11}v_1^2+\kappa_{12}v_1v_2+\kappa_{22}v_2^2 \label{eq:generaldragforcequadratic}
\ee
Where
\be
\frac{1}{\tau_{D1}} &=& \frac{\beta}{8\pi n_2}\intk{q}\int d\omega\intk{k_1}\,q_x(v_x(k_1)-v_x(k_1-q))\mathcal{A}(q)(f_0(k_1-q)-f_0(k_1)) \nonumber \\
                    &\times& \delta(\omega-\xi^t_{k_1}+\xi^t_{k_1-q})\frac{\Im \chi_2(q,\omega)}{\sinh^2(\beta\omega/2)} \label{eq:tauD1}
\ee
\be
\frac{1}{\tau_{D2}} &=& \frac{\beta}{8\pi n_2}\intk{q}\int d\omega\intk{k_2}\,q_x(v_x(k_2+q)-v_x(k_2))\mathcal{A}(q)(f_0(k_2)-f_0(k_2+q)) \nonumber \\
                    &\times& \delta(\omega+\xi^b_{k_2}-\xi^b_{k_2+q})\frac{\Im \chi_1(q,\omega)}{\sinh^2(\beta\omega/2)} \label{eq:tauD2} \\
                    &=& \frac{1}{m_b}\frac{\beta}{16\pi^2n_2}\intk{q}|q|^2\mathcal{A}(q)\int d\omega\,\frac{\Im\chi_1(q,\omega)\Im\chi_2(q,\omega)}{\sinh^2(\beta\omega/2)} \label{eq:tauD2simplified}
\ee
\be
\kappa_{11} &=& -\frac{(\beta\bar{m}^x_t)^2}{16\pi n_2}\intk{q}\int d\omega\,\intk{k_1}q_x\mathcal{A}(q)v_x(k_1-q)v_x(k_1)(f_0(k_1)-f_0(k_1-q))^2 \nonumber \\
            &\times& \delta(\omega-\xi^t_{k_1}+\xi^t_{k_1-q})\frac{\Im(\chi_2(q,\omega)-\chi_2(-q,-\omega))}{\sinh^2(\beta\omega/2)} \label{eq:kappa11} 
\ee
\be
\kappa_{12} &=& \frac{\beta^2 \bar{m}^x_t m_b}{2n_2}\intk{q}\intk{k_1}\intk{k_2}q_x\mathcal{A}(q)f_0(k_1)\tilde{f}_0(k_1-q)f_0(k_2)\tilde{f}_0(k_2+q)\delta(\xi^t_{k_1} \nonumber \\
            &+& \xi^b_{k_2}-\xi^t_{k_1-q}-\xi^b_{k_2+q})[v_x(k_1)v_x(k_2)(1-f_0(k_1)f_0(k_2))-v_x(k_1-q)v_x(k_2+q) \nonumber \\
            &\times& (1-f_0(k_1-q)f_0(k_2+q))+v_x(k_1)v_x(k_2+q)(f_0(k_1)-f_0(k_2+q)) \nonumber \\
            &-& v_x(k_1-q)v_x(k_2)(f_0(k_1-q)-f_0(k_2))] \label{eq:kappa12}
\ee
\be
\kappa_{22} &=& -\frac{(\beta m_b)^2}{16\pi n_2}\intk{q}\int d\omega\,\intk{k_2}q_x\mathcal{A}(q)v_x(k_2)v_x(k_2+q)(f_0(k_2)-f_0(k_2+q))^2 \nonumber \\
            &\times& \delta(\omega+\xi^b_{k_2}-\xi^b_{k_2+q})\frac{\Im(\chi_1(q,\omega)-\chi_1(-q,-\omega))}{\sinh^2(\beta\omega/2)} \label{eq:kappa22}
\ee
It is evident that when $\xi^t_k=\xi^t_{-k}$, $\kappa_{22}=0$; when $\xi^b_k=\xi^b_{-k}$, $\kappa_{11}=0$. When both previous symmetry conditions hold, $\kappa_{12}=0$ as well, hence all quadratic order terms vanish, forbidding nonreciprocal drag. In our case, $\kappa_{11}=0$, though the rest don't necessarily vanish. The non-vanishing of $\kappa_{12},\kappa_{22}$ leads to nonreciprocal Coulomb drag.
\newpara Combine \eqref{eq:generaldragforcequadratic} and \eqref{eq:forcebalance}, we obtain the following equation: 
\be
\frac{\bar{m}^x_t}{\tau_{D1}}v_1-\frac{m_b}{\tau_{D2}}v_2+\kappa_{11}v_1^2+\kappa_{12}v_1v_2+\kappa_{22}v_2^2-\frac{m_b}{\tau_2}v_2 = 0 \label{eq:quadraticforcebalance}
\ee
Where $\kappa_{11}=0$ is our case. Noticing the typical smallness of $v_2/v_1$ in frictional drag regime, we keep nonlinear terms in the equation to the leading order of $v_2/v_1$ and obtain
\be
\frac{v_2}{v_1} &=& \frac{\bar{m}^x_t/\tau_{D1}+\kappa_{11}v_1+\kappa_{12}v_2}{m_b(1/\tau_2+1/\tau_{D2})+\kappa_{22}v_2} \\
                &\approx& \frac{\bar{m}^x_t}{m_b(\tau_{D1}/\tau_2+\tau_{D1}/\tau_{D2})}+(\frac{\kappa_{12}\tau_{D1}}{m_b}-\frac{\kappa_{22}\tau_{D1}(\bar{m}^x_t+\kappa_{11}\tau_{D1}v_1)}{m_b^2})v_2 \\
                &\approx& \frac{\bar{m}^x_t}{m_b(\tau_{D1}/\tau_2+\tau_{D1}/\tau_{D2})}[1+(\frac{\kappa_{12}\tau_{D1}}{m_b}-\frac{\kappa_{22}\tau_{D1}\bar{m}^x_t}{m_b^2})v_1] \label{eq:v2/v1quadratic}
\ee
Namely, up to next leading order, drag ratio $\zeta=\zeta^{(0)}+\zeta^{(1)}(v_1)$ is a function of $v_1$, where
\be
\zeta^{(0)} &=& \frac{n_2e_2}{n_1e_1}\frac{\bar{m}^x_t}{m_b(\tau_{D1}/\tau_2+\tau_{D1}/\tau_{D2})} \label{eq:zeta0} \\
\zeta^{(1)}(v_1) &=& \zeta^{(0)}(\frac{\kappa_{12}\tau_{D1}}{m_b}-\frac{\kappa_{22}\tau_{D1}\bar{m}^x_t}{m_b^2})v_1 \label{eq:zeta1}
\ee
The nonzero $\zeta^{(1)}(v_1)$ term is responsible for both nonlinear and nonreciprocal drag effects.

\subsubsection{Application to electron-hole bilayer}

In the electron-hole bilayer setup where $\xi^{t(b)}_k=\xi_{e(h),k}=\frac{k^2}{2m_{e(h)}}-\mu_{e(h)} - |\lambda_{e(h)}|\sqrt{(k_x+\frac{B}{\lambda_{e(h)}})^2+k_y^2+(\frac{J}{\lambda_{e(h)}})^2}$, which is extensively discussed in the main text. We have
\be
\zeta = \frac{I_{\text{drag}}}{I_{\text{drive}}} = \zeta^{(0)}+\zeta^{(1)}(j_{drive})
\ee
Note $I_{\text{drive,drag}}=j_{\text{drive,drag}}\cdot L_{y}$, where $L_y$ is the effective length of the sample in \y direction, which plays the role of the 1-dimensional cross section of a 2-dimensional sample.
\be
\zeta^{(0)} &=& -\frac{n_h}{n_e}\frac{\bar{m}_e^x}{m_h}\frac{1}{\tau_{D1}/\tau_b+\tau_{D1}/\tau_{D2}}\\
\zeta^{(1)} &=& -\zeta^{(0)}(\frac{\kappa_{12}\tau_{D1}}{m_h}-\frac{\kappa_{22}\tau_{D1}\bar{m}_e^x}{m_h^2})\frac{j_{\text{drive}}}{n_e e}
\ee
Where now
\be
\frac{1}{\tau_{D1}} &=& \frac{\beta}{8\pi n_h}\intk{q}\int d\omega\intk{k}\,q_x(v^e_x(k)-v^e_x(k-q))\mathcal{A}(q)(f_0(\xi_{e,k-q})-f_0(\xi_{e,k})) \nonumber \\
                    &\times& \delta(\omega-\xi_{e,k_1}+\xi_{e,k_1-q})\frac{\Im \chi_h(q,\omega)}{\sinh^2(\beta\omega/2)} 
\ee
\be
\frac{1}{\tau_{D2}} &=& \frac{\beta}{8\pi n_h}\intk{q}\int d\omega\intk{k}\,q_x(v^h_x(k+q)-v^h_x(k))\mathcal{A}(q)(f_0(\xi_{h,k})-f_0(\xi_{h,k+q})) \nonumber \\
                    &\times& \delta(\omega+\xi_{h,k}-\xi_{h,k+q})\frac{\Im \chi_e(q,\omega)}{\sinh^2(\beta\omega/2)} \\
                    &=& \frac{1}{m_h}\frac{\beta}{16\pi^2n_h}\intk{q}|q|^2\mathcal{A}(q)\int d\omega\,\frac{\Im\chi_e(q,\omega)\Im\chi_h(q,\omega)}{\sinh^2(\beta\omega/2)} 
\ee
\be
\kappa_{12} &=& \frac{\beta^2 \bar{m}^x_e m_h}{2n_h}\intk{q}\intk{k_1}\intk{k_2}q_x\mathcal{A}(q)f_0(\xi_{e,k_1})\tilde{f}_0(\xi_{e,k_1-q})f_0(\xi_{h,k_2})\tilde{f}_0(\xi_{h,k_2+q})\delta(\xi_{e,k_1} \nonumber \\
            &+& \xi_{h,k_2}-\xi_{e,k_1-q}-\xi_{h,k_2+q})[v_x^e(k_1)v_x^h(k_2)(1-f_0(\xi_{e,k_1})f_0(\xi_{h,k_2}))-v_x^e(k_1-q)v_x^h(k_2+q) \nonumber \\
            &\times& (1-f_0(\xi_{e,k_1-q})f_0(\xi_{h,k_2+q}))+v_x^e(k_1)v_x^h(k_2+q)(f_0(\xi_{e,k_1})-f_0(\xi_{h,k_2+q})) \nonumber \\
            &-& v_x^e(k_1-q)v_x^h(k_2)(f_0(\xi_{e,k_1-q})-f_0(\xi_{h,k_2}))] 
\ee
\be
\kappa_{22} &=& -\frac{(\beta m_h)^2}{16\pi n_h}\intk{q}\int d\omega\,\intk{k}q_x\mathcal{A}(q)v_x^h(k)v_x^h(k_2)(f_0(\xi_{h,k})-f_0(\xi_{h,k+q}))^2 \nonumber \\
            &\times& \delta(\omega+\xi_{h,k}-\xi_{h,k+q})\frac{\Im(\chi_e(q,\omega)-\chi_e(-q,-\omega))}{\sinh^2(\beta\omega/2)}  
\ee
As what has been discussed previously, nonreciprocal frictional Coulomb drag comes from $\zeta^{(1)}$ term. Yet when $\xi_{e,k}=\xi_{e,-k}$, both $\kappa_{12}$ and $\kappa_{22}$ will vanish, which will make $\zeta^{(1)}$ term vanish. If $\xi_{e,k}\neq\xi_{e,-k}$ as in our model, nonreciprocal frictional drag is present in general.

\bibliography{XSDE.bib}